\documentclass[11pt]{article}

\usepackage{latexsym,amsfonts,amsmath,amssymb,wasysym,enumerate,epsfig}
\usepackage{theorem}
\usepackage{mathrsfs}
\usepackage{tikz}
\usepackage[all]{xy}

\usepackage{color}

\numberwithin{equation}{section}

\newcommand{\mb}[1]{\quad\mbox{#1}\quad}
\newcommand{\be}{\begin{equation}}
\newcommand{\ee}{\end{equation}}
\newcommand{\beu}{\begin{equation*}}
\newcommand{\eeu}{\end{equation*}}
\newcommand{\bea}{\begin{eqnarray}}
\newcommand{\eea}{\end{eqnarray}}
\newcommand{\beaa}{\begin{eqnarray*}}
\newcommand{\eeaa}{\end{eqnarray*}}
\newcommand{\bmx}{\begin{pmatrix}}
\newcommand{\emx}{\end{pmatrix}}

\newcommand{\D}{{\cal D}}

\newcommand{\half}{\frac{1}{2}}

\newcommand{\eps}{\epsilon}

\newcommand{\CC}{{\mathbb C}}

\def\cA{{\cal A}}        
\def\cD{{\cal D}}

\newcommand{\atopn}[2]{\genfrac{}{}{0pt}{}{#1}{#2}}

\newcommand{\llangle}{\langle\!\langle}
\newcommand{\rrangle}{\rangle\!\rangle}
\newcommand{\steady}{|{\cal S}\rangle} 
\advance\textheight by \topskip
\begin{document}
\setcounter{page}{0}
\pagestyle{empty}
%
%
\begin{center}

 {\LARGE  {\sffamily Integrable approach to simple exclusion processes with boundaries.\\[1.2ex]
Review and progress} }\\[1cm]

\vspace{10mm}
  
{\Large 
 N. Crampe$^{ab,}$\footnote{nicolas.crampe@univ-montp2.fr},
 E. Ragoucy$^{c}$\footnote{eric.ragoucy@lapth.cnrs.fr}
 and M. Vanicat$^{c}$\footnote{matthieu.vanicat@lapth.cnrs.fr}}\\[.41cm] 
{\large $^a$ Universit\'e Montpellier 2, Laboratoire Charles Coulomb UMR 5221,\\[.242cm]
F-34095 Montpellier, France}
\\[.42cm]
{\large $^b$ CNRS, Laboratoire Charles Coulomb, UMR 5221, F-34095 Montpellier}
\\[.42cm]
{\large $^{c}$ Laboratoire de Physique Th{\'e}orique LAPTh
 CNRS and Universit{\'e} de Savoie.\\[.242cm]
   9 chemin de Bellevue, BP 110, F-74941  Annecy-le-Vieux Cedex, 
France. }
\end{center}
\vfill

\begin{abstract}
We study the matrix ansatz in the quantum group framework, applying integrable systems techniques to statistical physics models. 
We start by reviewing the two approaches, and then show how one can use the former to get new insight on the latter. 
We illustrate our method by solving a model of reaction-diffusion.
An eigenvector for the transfer matrix for the XXZ spin chain with non-diagonal boundary is also obtained using a matrix 
ansatz.
\end{abstract}

\vfill\vfill
\rightline{LAPTh-074/14}
\rightline{August 2014}

\newpage
\pagestyle{plain}
\section*{Introduction}

In statistical physics, one-dimensional models of particles with hard-core interactions and nearest-neighbor
hopping, called Asymmetric Simple Exclusion Process (ASEP), provide basic example of out-of-equilibrium systems.
These models, despite their simplicity, behave non trivially and exhibit interesting 
physical phenomena such as phase transitions induced by the boundaries \cite{krug}.
The matrix ansatz (or DEHP method, named according to B. Derrida, M. Evans, V. Hakim and V. Pasquier) has been developed to 
find their stationary state in \cite{DEHP,sandow} and to obtain analytically the phase diagram in term 
of the injection/extraction rates at the boundaries. This method has been
intensively studied and generalized for different models (see e.g. \cite{BE} for a review).

In the context of integrable models, the notion of integrable boundaries has been introduced in \cite{sklyanin}.
It was a starting point for the development of the algebraic Bethe ansatz that allows one
to compute the eigenvectors and eigenvalues of integrable models with boundaries and, in particular, 
of the XXZ spin chain (see section \ref{sec:ba} for a state of the art).
It is now well established that the underlying mathematical structure for integrability are Hopf algebras and their coideals. 
It led to the development of the so-called quantum groups and reflection algebras.

In this paper, we intend to study the relations between quantum groups and matrix ansatz, 
connecting a statistical physics approach to  integrable techniques.
Of course, the first step consists in proving that a given out-of-equilibrium model is integrable:
one must find its $R$-matrix, solution to the Yang-Baxter equation, and the $K$-matrix, solution to the reflection
equation. They allow one to construct a transfer matrix that commutes with the Markov matrix \cite{sklyanin}.
This $R$-matrix can be also used to define a quantum group, thanks to the so-called FRT relation \cite{FRT} 
(named according to its authors L. Faddeev, N. Reshetikhin and L. Takhtajan).
Then, using a supplementary property of the $R$-matrix, called Markovian property 
\eqref{eq:marmr}, we succeed in proving that the so-called Zamolodchikov algebra is a subalgebra of this quantum group. 
Let us emphasize that this point was not known previously since the $R$-matrix associated to a spin chain does not
satisfy in general the Markovian property. Finally, from the Zamolodchikov algebra, one gets the bulk part of the 
matrix ansatz. This connection has been already noticed in \cite{SW}. 
As a consequence, we can express the generators of the matrix ansatz in terms of quantum group generators. 
These connections are summarized in figure \ref{fig:mc}.
\begin{figure}[htb]
\begin{center}
 \begin{tikzpicture}[scale=0.7]
 \node at (-8,-3) [] {Quantum group};
 \node at (-8,-3.8) [] {$R_{12}T_1T_2=T_2T_1R_{12} $};
 \draw[->] (-5,-3.4) -- (-3,-3.4);
 \node at (0,-3) [] {Zamolodchikov algebra};
 \node at (0,-3.8) [] {$R_{12}A_1A_2=A_2A_1$};
 \draw[->] (3,-3.4) -- (5,-3.4);
 \node at (8,-3) [] {Matrix ansatz};
 \node at (8,-3.8) [] {$w_{12}s_1s_2=s_1\bar s_2-\bar s_1s_2$};
 \end{tikzpicture}
 \vspace{-3ex}
 \end{center}
 \caption{Main connections presented in this paper (bulk part).}
 \label{fig:mc}
\end{figure}

For the boundary part of the matrix ansatz, the same connection exists, but it is based on the Ghoshal-Zamolodchikov relations, defined with the $K$-matrix, see figure \ref{fig:gz}.
This relation has been introduced previously in the context of  
integrable quantum field with boundaries \cite{GZ}. These relations allows us to compute one eigenvector and eigenvalue of the inhomogeneous transfer matrix,
which is a complicate task for non-diagonal boundaries, see e.g. \cite{CYSW03,YZ,FKN,KMN}.
\begin{figure}[htb]
\begin{center}
 \begin{tikzpicture}[scale=0.7]
 \node at (-8,-3) [] {Reflection equation};
 \node at (-8,-3.8) [] {$R_{12}K_1R'_{21}K_2=K_2R'_{12}K_1R_{21} $};
 \draw[->] (-5,-3.4) -- (-3,-3.4);
 \node at (0,-3) [] {G.Z. relation};
 \node at (0,-3.8) [] {$\llangle\Omega | K(x) A(\frac1x)= \llangle\Omega |A(x) $};
 \draw[->] (3,-3.4) -- (5,-3.4);
 \node at (8,-3) [] {Boundary condition};
 \node at (8,-3.8) [] {$\llangle W|{B}\,s=\llangle W|\bar s$};
 \end{tikzpicture}
 \vspace{-3ex}
 \end{center}
 \caption{Main connections presented in this paper (boundary part).}.
 \label{fig:gz}
 \vspace{-3ex}
\end{figure}

We want also to stress that the computation of the stationary state is always possible by a matrix ansatz.
However this computation is, in general, as hard as a direct diagonalization of the Markov process \cite{KS}.
The links presented here between the matrix ansatz and some algebras coming from integrability framework
allow us to facilitate this calculation when the models are integrable. Therefore, our framework  
relates the notion of integrability of the model to an efficient matrix ansatz.
To illustrate this point, we present in this paper the construction of the stationary state of an integrable reaction-diffusion model. 

It is well-known that ASEP and XXZ models
are related  \cite{sandow}. However, the two models are paradigmatic for two different disciplines, statistical physics (for ASEP) 
and integrable models (for XXZ). Thus, when dealing with these two equivalent archetypes, the technique and languages are nevertheless different. 
Hence, we need to detail the two view points on these models, dealing with very well-known concepts 
for at least one of the two communities. Yet, we believe it is a necessary 
path to connect the two domains.
It is the motivation for our section \ref{sec:mark} that reviews well-known notions concerning one-dimensional out-of-equilibrium system:
we discuss general notions concerning the Markov processes, describe more explicitly the
exclusion processes used in each section to illustrate the different notions and present the usual matrix ansatz. 
Section \ref{sec:integrability} is the mirror image of section \ref{sec:mark} for integrable models. We introduce the $R$-matrix, solution to the Yang-Baxter equation 
and the $K$-matrix, solution to the reflection equation. They are the fundamental objects needed
to construct the transfer matrix, allowing one to prove the integrability of these models. Then, in section \ref{quantum_group}, the more algebraic one, 
the notion of quantum group is recalled briefly and, in particular, its FRT formalism using the $R$-matrix. We present also the Zamolodchikov algebra,
its link with the FRT formalism and the Ghoshal-Zamolodchikov relations. 
The section \ref{sec:link} is the heart of the paper: we prove that 
the Zamolodchikov algebra and the Ghoshal-Zamolodchikov relations generalize the matrix ansatz. 
The two last sections concern the use of this generalization. We compute the eigenvector and eigenvalue
corresponding to the steady state of the transfer matrix (section \ref{sec:ti}) 
and compare it with the Bethe ansatz approaches. We compute the stationary state, densities and currents of a reaction-diffusion model (section \ref{sec:8v}).

\section{Markov matrix and matrix ansatz\label{sec:mark}}

We review here the notion of a one-dimensional Markov process and its associated master equation
as well as the matrix ansatz allowing one to compute its stationary state.

\subsection{Markov process\label{sec:MP}}

Let us consider a one-dimensional lattice with $L$ sites where each site can be empty or occupied by a particle.
Hence, a configuration of the system is a $L$-tuple $\mathcal{C}=(\tau_1,\tau_2,\dots,\tau_L)$ 
where $\tau_i=0$ if the site $i$ is empty and $\tau_i=1$ if a particle\footnote{Evidently, one can generalize to problems where many particles can be 
simultaneously on the same site, or to particles with internal degrees of freedom, or with different types of particles. 
To simplify  this overview, we restrict ourselves to case with hard core exclusion, without internal degrees of freedom and one type of particle.}
 is on  site $i$. 
The dynamics of the particles is stochastic and $M(\mathcal{C},\mathcal{C}')$ denotes the transition rate from the configuration $\mathcal{C}'$ to 
$\mathcal{C}$: if the system is in the state $\mathcal{C}'$, it has a probability $M(\mathcal{C},\mathcal{C}')\times \delta t$ to jump 
to the state $\mathcal{C}$ during the time interval $\delta t$. One considers only Markovian models \textit{i.e.} the rates $M(\mathcal{C},\mathcal{C}')$
depends only on $\mathcal{C}$ and $\mathcal{C}'$ and not on the history of the system.
If $P_t(\mathcal{C})$ is the probability to find the system in the configuration $\mathcal{C}$ at time $t$, its time evolution 
is then governed by the following master equation 
\begin{equation}\label{eq:ME}
 \frac{d P_t(\mathcal{C})}{dt}=\sum_{\mathcal{C}'\neq \mathcal{C}}M(\mathcal{C},\mathcal{C}')P_t(\mathcal{C}')
 -\sum_{\mathcal{C}'\neq \mathcal{C}}M(\mathcal{C}',\mathcal{C}) P_t(\mathcal{C})\;.
\end{equation}

A useful presentation of the previous equation consists in gathering all the probabilities in a vector as follows
\begin{equation}
 |P_t\rangle=\left(
 \begin{array}{c}
  P_t(\ (0,\dots,0,0,0)\ )\\
   P_t(\ (0,\dots,0,0,1)\ )\\
   P_t(\ (0,\dots,0,1,0)\ )\\
   \vdots\\
   P_t(\ (1,\dots,1,1,1)\ )
 \end{array}
\right)
=\sum_{\tau_1,\dots,\tau_L\in \{0,1\}}  P_t(\ (\tau_1,\dots,\tau_{L})\ )~ e^{\tau_1}\otimes\dots\otimes e^{\tau_L}
\end{equation}
where $e^{0}=\left(
 \begin{array}{c}
 1\\0
 \end{array}
\right)$ and $e^{1}=\left(
 \begin{array}{c}
 0\\1
 \end{array}
\right)$. From the previous definition, it is clear that the set of all configurations can be seen as the tensor space  
$\left(\mathbb{C}^2\right)^{\otimes L}=\underbrace{\mathbb{C}^2\otimes\dots\otimes\mathbb{C}^2}_{L \text{ times}}$ where the $j^{\text{th}}$ space
$\mathbb{C}^2$ corresponds to the $j^{\text{th}}$ site of the lattice. This space is called the configuration space.
We define also $M(\mathcal{C},\mathcal{C})=-\sum_{\mathcal{C}'\neq \mathcal{C}}M(\mathcal{C}',\mathcal{C})$ which allows us to introduce 
the Markov matrix $M$ 
whose entries are $M(\mathcal{C},\mathcal{C}')$.
With these notations, the master equation \eqref{eq:ME} can be rewritten in a more concise way
\begin{equation}\label{eq:MEv}
 \frac{d |P_t\rangle}{dt}=M\ |P_t\rangle.
\end{equation}

In addition, we assume that the configuration
can change only between two adjacent sites at the same time and that there is a translation invariance in the bulk \textit{i.e.} 
the rate of this change is independent of the site. 
We assume also that the boundaries (the sites $1$ and $L$) are in contact with reservoirs and that supplementary changes of configuration 
can take place at these sites.
With these hypotheses, the Markov matrix $M$ can be written in terms of a sum of local jump operators as follows
\bea \label{eq:W}
M &=&  B_1 + \sum_{\ell=1}^{L-1} w_{\ell,\ell+1} + \overline{B}_L
\eea
where the subscribed index ($1$, $L$ or $\ell,\ell+1$) in \eqref{eq:W} indicate on which component of the tensor space $\left(\mathbb{C}^2\right)^{\otimes L}$ 
the local operators $B$, $\overline{B}$ and $w$ act non trivially.

\subsection{Example: the Asymmetric Simple Exclusion Process\label{sec:ASEP}}

To illustrate the different concepts discussed in this paper, we use well-known Markov process
called the Asymmetric Simple Exclusion Process (ASEP) with open boundaries 
as well as two different limits: the Symmetric Simple Exclusion Process (SSEP) and the Totally Asymmetric Simple Exclusion Process (TASEP).
We introduce another example in section \ref{sec:8v}.

\paragraph{ASEP model.}
The dynamics of the ASEP model can be summarized by the following rules
\begin{eqnarray}
\begin{array}{ c c c}
\text{left boundary}\hspace{1cm}&\text{bulk}\hspace{1cm}&\text{right boundary}\\
 0 \xrightarrow{\hspace{2mm}\alpha\hspace{2mm}} 1\hspace{1cm} & 01 \xrightarrow{\hspace{2mm}q\hspace{2mm}} 10\hspace{1cm}& 0 \xrightarrow{\hspace{2mm}\delta\hspace{2mm}} 1\\
 1 \xrightarrow{\hspace{2mm}\gamma\hspace{2mm}} 0\hspace{1cm}& 10 \xrightarrow{\hspace{2mm}1\hspace{2mm}} 01\hspace{1cm}&  1 \xrightarrow{\hspace{2mm}\beta\hspace{2mm}} 0
 \end{array}
\end{eqnarray}
For this process, the local jump operators are given explicitly by
\begin{equation}
 B =\left( \begin {array}{cc} 
-\alpha&\gamma\\ 
\alpha&-\gamma
\end {array} \right)\quad,\qquad   
w=\left( \begin {array}{cccc} 
0&0&0&0\\ 
0&-q&1&0\\
0&q&-1&0\\
0&0&0&0
\end {array} \right) \quad,\qquad 
\overline{B} =\left( \begin {array}{cc} 
-\delta&\beta\\ 
\delta&-\beta
\end {array} \right)\:.
\label{eq:localwASEP}
\end{equation}
This explicit form of $M$ allows one to link this model with the quantum spin chain called XXZ \cite{sandow}. 
Indeed, by a gauge transformation (not necessarily unitary), one can map the Markov matrix $M$ to the Hamiltonian of the XXZ model. 

\paragraph{SSEP model ($q=1$).}
The limit $q\to1$ gives the SSEP model where
\begin{equation}\label{eq:ws}
w^{SSEP}=\left( \begin {array}{cccc} 
0&0&0&0\\ 
0&-1&1&0\\
0&1&-1&0\\
0&0&0&0
\end {array} \right), 
\end{equation}
and the boundary matrices are still given by (\ref{eq:localwASEP}).
In this limit, we have a vanishing external field: in the bulk the particles can go to the left or to the right with the same probability.
The system is put out-of-equilibrium only by the reservoirs at the two boundaries. This model is linked to the quantum spin chain 
called XXX or Heisenberg model. At first glance, the limit $q\rightarrow 1$ seems trivial. However, in numerous cases, it requires 
a careful rescaling of the parameters before performing the limit. Therefore, this model is often studied on its own.

\paragraph{TASEP model ($q=0$).}

The limit $q\to0$ gives the TASEP model where
\begin{equation}\label{eq:wta}
B^{TASEP} =\left( \begin {array}{cc} 
-\alpha&0\\ 
\alpha&0
\end {array} \right)\quad,\quad w^{TASEP}=\left( \begin {array}{cccc} 
0&0&0&0\\ 
0&0&1&0\\
0&0&-1&0\\
0&0&0&0
\end {array} \right)\quad,\quad
\overline{B}^{TASEP} =\left( \begin {array}{cc} 
0&\beta\\ 
0&-\beta
\end {array} \right) \:.
\end{equation}
In this case the particles can only move to the right, the jumps to the left being now forbidden. The gauge transformation to get a quantum Hamiltonian 
related to a limit of the XXZ model is not invertible anymore. Therefore, there is no quantum Hermitian Hamiltonian associated to this model. 

\subsection{Matrix ansatz\label{matrix_ansatz}}

In this section, we recall the matrix ansatz (or DEHP method) introduced in \cite{DEHP} (see also \cite{BE} for a review) to 
compute the stationary state of the process \textit{i.e.} the probability distribution which satisfies
\begin{equation}
\frac{d {\cal S}(\mathcal{C})}{dt}=0=
\sum_{\mathcal{C}'\neq \mathcal{C}}M(\mathcal{C},\mathcal{C}'){\cal S}(\mathcal{C}')
-\sum_{\mathcal{C}'\neq \mathcal{C}}M(\mathcal{C}',\mathcal{C}) {\cal S}(\mathcal{C})
\end{equation}
where ${\cal S}({\cal C})$ is the probability for the system to be in the configuration $\cal C$ for the stationary state.
Gathering all the probabilities ${\cal S}({\cal C})$ in the vector $\steady$, one sees that 
the stationary property is equivalent to find an eigenvector of $M$ with 
vanishing eigenvalue:  $M\steady=0$.

Before presenting the matrix ansatz, let us first recall few information concerning the stationary state. The Perron-Frobenius theorem tells us that this stationary 
state is unique since the Markov matrix $M$ has a non degenerate zero eigenvalue  
(because the process is irreducible: one can go from any configuration $\mathcal{C}$ to any other configuration $\mathcal{C}'$ in a finite number
of jumps with a non zero probability). Moreover all the other eigenvalues 
of $M$ have a negative real part, so that the probability vector $|P_t\rangle$ will converge (for $t\to\infty$) to the stationary state $\steady$.
Let us also remark that the left eigenvector corresponding to $\steady$ is easy to compute: all the column of $M$ sum up to $0$,
so the vector $\langle 1|$ with all entries $1$ is the left eigenvector associated to the $0$ eigenvalue: $\langle 1|M=0$. 

\subsubsection{The general ansatz}

The matrix ansatz consists in looking for the probability of the  stationary state configuration
$\mathcal{C}=(\tau_1,\tau_2,\dots,\tau_L)$  in the following form
\begin{equation}\label{eq:sc}
  {\cal S}({\cal C})=\frac{1}{Z_L}\ \llangle W| \prod_{i=1}^{L} \big(\ (1-\tau_i)E+\tau_i D\ \big) |V\rrangle,
\end{equation}
where $E$ and $D$ are elements of a non commutative algebra $\cA$ and the product must be understood as ordered from $1$ on the left to $L$ on the right.
The factor $Z_L$ is a normalization such that $\sum_{\cal C} {\cal S}({\cal C})=1$ and can be written as follows
\begin{equation}
 Z_L= \llangle W| (E+D)^L |V\rrangle\;.
\end{equation}

At this stage, let us stress that $|V\rrangle$ and $\llangle W|$ belong to an additional space that 
is different from the configuration space $(\CC^2)^{\otimes L}$. This additional space is in fact a representation space for the algebra $\cA$. To make this distinction clear enough, we use different notations for vectors belonging to these two spaces:
 $|.\rangle$ stands for vectors in the configuration space while $|.\rrangle$ refers to vectors in the additional 
 space used for the matrix ansatz. The additional 
 space is sometimes called auxiliary space but, in this paper, this term will be used for another type of space when we will consider 
 the so-called transfer matrix coming from integrable systems framework.

The commutation relation between $E$ and $D$
and their action on $\llangle W|$ and $|V\rrangle$ encode all the information needed to describe the steady state $\steady$ 
and depend on the considered model.
Indeed, in \cite{KS}, it was shown that for the Markov matrices of type \eqref{eq:W}, it is always possible to construct the stationary state 
with the ansatz \eqref{eq:sc} where $E$, $D$, $|V\rrangle$ and $\llangle W|$ satisfy
\begin{eqnarray}\label{eq:tele}
&& w\left(\begin{array}{c}E\\D
 \end{array}\right)\otimes \left(\begin{array}{c}E\\D
 \end{array}\right)=\left(\begin{array}{c}E\\D
 \end{array}\right)\otimes \left(\begin{array}{c}\overline{E}\\\overline{D}
 \end{array}\right)-\left(\begin{array}{c}\overline{E}\\\overline{D}
 \end{array}\right)\otimes \left(\begin{array}{c}E\\D
 \end{array}\right)\;,\\
 \label{eq:telB}
 &&\llangle W| B\left(\begin{array}{c}E\\D
 \end{array}\right)=\llangle W|\left(\begin{array}{c}\overline{E}\\ \overline{D}
 \end{array}\right)\quad\text{and}\qquad
 \overline{B}\left(\begin{array}{c}E\\D
 \end{array}\right)|V\rrangle=-\left(\begin{array}{c}\overline{E}\\ \overline{D}
 \end{array}\right)|V\rrangle
\end{eqnarray}
and $\overline{E}$, $\overline{D}$ are in general two new non-commuting elements.
It is easy to show that $\steady$ given by \eqref{eq:sc}, when \eqref{eq:tele} and \eqref{eq:telB} hold, satisfies $M\steady=0$ 
(see section \ref{sec:notation}). The difficult part of the theorem in \cite{KS} consisted in proving that $\steady\neq 0$.\\

The matrix ansatz is often used when the operators $\overline{E}$, $\overline{D}$ are scalars because it leads to simpler calculations.
It was done for the TASEP and ASEP models with open boundary \cite{DEHP,sandow}, where explicit representations were constructed for the $E$, $D$
operators  and the vectors $|V\rrangle$ and $\llangle W|$. 
We give a brief review of these results in section \ref{sec:MAex}.
A classification of the models solvable by matrix ansatz with scalar $\overline{E}$ and $\overline{D}$,
 and the study of the corresponding quadratic algebras, was done in \cite{IPR}. In particular it was shown that the associativity of the algebra put
strong constraints on the parameters of the model.
In some cases, non scalar operators $\overline{E}$ and $\overline{D}$ are necessary. For example,
a reaction-diffusion model has been solved with this method in \cite{HSP}. 
More recently a matrix ansatz involving non scalar overlined operators for the multispecies TASEP and ASEP models on the ring was built recursively in \cite{PEM,EFM,AAMP}.
In \cite{HS}, an explicit construction using finite matrices of these operators for small system size is proposed.\\
 
As explained previously, the construction of the stationary state from the matrix ansatz is always possible 
but may be as hard as a brute force diagonalization of $M$ \cite{KS}. 
One goal of this paper is to provide a general framework
where, once the integrability of the model is established, we are able to construct an efficient matrix ansatz.
Indeed the relations \eqref{eq:tele} and \eqref{eq:telB} are not sufficient to compute any word in $E$ and $D$ when 
$\overline{E}$ and $\overline{D}$ are not scalars. The framework proposed in this paper (in particular  
the Zamolodchikov algebra and the Ghoshal-Zamolodchikov relations) allows us to obtain the necessary and sufficient 
conditions between the generators of the algebra to compute all the words.

\subsubsection{Proof of the stationary state \label{sec:notation}}

We give here the well-known proof that $M\steady=0$. We use notations borrowed from the 
quantum inverse scattering method \cite{QISM}
that allow one to treat objects (vectors, matrices,...) living in different tensor spaces. It allows us to illustrate
these notations and show how powerful they are.
Let us introduce
\begin{equation}
 s=\left(\begin{array}{c}E\\D
 \end{array}\right)\quad\mbox{and}\qquad \bar s=\left(\begin{array}{c}\overline{E}\\ \overline{D}
 \end{array}\right)\:.
\end{equation}
Remark that $s$ and $\bar s$ belong to the tensor product $\CC^2\otimes\cA$ and should be noted more precisely as, e.g. $s= e^0 \otimes E+ e^1\otimes D$.
The stationary state can be written as follows
\begin{equation}\label{eq:O}
 \steady=\frac{1}{Z_L}\ \left(
 \begin{array}{c}
 \llangle W|E\dots EEE|V\rrangle\\
  \llangle W|E\dots EED|V\rrangle\\
 \llangle W|E\dots EDE|V\rrangle\\
   \vdots\\
   \llangle W|D\dots DDD|V\rrangle
 \end{array}
\right)=\frac{1}{Z_L}\ \llangle W|~\prod_{j=1}^L s_{j}~ |V\rrangle\;,
\end{equation} 
where the subscript $j$ on $s$ indicates the $\CC^2$-space (among $(\CC^2)^{\otimes L}$) the corresponding $s$ belongs to. Note that the entries of all the $s_j$ belong to the same algebra $\cA$.
Hence $\prod_{j=1}^L s_{j}$ belongs to $(\CC^2)^{\otimes L}\otimes\cA$, and the action of the vectors $|V\rrangle$ and $\llangle W|$ (that define a representation of $\cA$) makes $\steady$ a vector of $(\CC^2)^{\otimes L}$ only.
Let us emphasize that the entries of $s$ are not commuting and that the product must be understood as ordered from $1$ to $L$. With these notations,  relations \eqref{eq:tele} and \eqref{eq:telB} are recast as
\bea
\label{eq:tele-aux}
&& w_{12}\,s_1\,s_2=s_1\,\bar s_2-\bar s_1\,s_2\;,\quad
\llangle W|\, B\,s=\llangle W|\,\bar s\quad\text{and}\quad
 \overline{B}\,s\,|V\rrangle=-\bar s\,|V\rrangle\,.
\end{eqnarray}

Then, the proof of $M\steady=0$ is given by
\begin{eqnarray}
 Z_L\ M\steady&=&M ~\llangle W|~\prod_{j=1}^L s_{j}~ |V\rrangle\label{eq:prg}  = \llangle W|~M~\prod_{j=1}^L s_{j}~ |V\rrangle \label{eq:preuve1}\\
 &=& \llangle W|~\left(B_1 + \sum_{\ell=1}^{L-1} w_{\ell,\ell+1} + \overline{B}_L\right)~\prod_{j=1}^L s_{j}~ |V\rrangle\label{eq:preuve1bis}\\ 
 &=& \llangle W|~\left(B_{1}s_{1}\right)s_{2}\dots s_{L}~ |V\rrangle+\llangle W|~s_{1}\dots s_{L-1} \left(\overline{B}_{L} s_{L}\right)~ |V\rrangle \label{eq:preuve2}\\
 & & + \sum_{\ell=1}^{L-1}\llangle W|~s_{1} \dots s_{\ell-1}\left( w_{\ell,\ell+1}s_{\ell}s_{\ell+1}\right)s_{\ell+2}\dots s_{L}~ |V\rrangle\nonumber\\
 &=& \llangle W|~\bar s_{1}s_{2}\dots s_{L}~ |V\rrangle-\llangle W|~s_{1}\dots s_{L-1} \bar s_{L}~ |V\rrangle \\
 & & + \sum_{\ell=1}^{L-1}\llangle W|~s_{1} \dots s_{\ell-1}\left(s_{\ell}\bar s_{\ell+1}- \bar s_{\ell}s_{\ell+1}\right)s_{\ell+2}\dots s_{L}~ |V\rrangle\nonumber\\
 &=& 0.
\end{eqnarray}
The second equality in \eqref{eq:preuve1} follows from the fact that the Markov matrix $M$ acts on the configuration space only, independently from the vector $\llangle W|$
which lies in the additional space. Then one uses the explicit form \eqref{eq:W} of $M$ to get \eqref{eq:preuve1bis}. The equality \eqref{eq:preuve2}
is obtained by moving $\overline{B}_{L}$ through all the $s_{1}$,..., $s_{L-1}$ operators (respectively $w_{\ell,\ell+1}$ through the $s_{1}$,..., $s_{\ell-1}$ operators)
because it acts trivially on these components of the tensor space. Using \eqref{eq:tele-aux}, one finally gets a telescopic sum which vanishes.

\subsubsection{Examples\label{sec:MAex}}

In this section, we give the different matrix ansatz for the different Markov processes introduced in Section \ref{sec:ASEP}.
In this case, it can be simplified tremendously because the 
elements $\overline{E}$ and $\overline{D}$ can be chosen proportional to identity.

\paragraph{ASEP model.}
$E$ and $D$ are elements of a quadratic algebra $\cA$ defined by the following relation
\begin{eqnarray}
 \label{eq:tel}
 w  \left(\begin{array}{c}E\\D
 \end{array}\right) \otimes \left(\begin{array}{c}E\\D
 \end{array}\right) =(q-1) \left[ \left(\begin{array}{c}E\\D
 \end{array}\right) \otimes \left(\begin{array}{c}1\\-1
 \end{array}\right)-\left(\begin{array}{c}1\\-1
 \end{array}\right) \otimes  \left(\begin{array}{c}E\\D
 \end{array}\right)\right]\;,
\end{eqnarray}
where $w$ is given by \eqref{eq:localwASEP} and we chose $\overline{E}=q-1$ and $\overline{D}=1-q$.
Equality \eqref{eq:tel} is written in $\CC^2\otimes\CC^2\otimes\cA$: once projected on a basis of $\CC^2\otimes\CC^2$, it gives four relations.
Writing the four components of \eqref{eq:tel}, one can show that they are in fact equivalent to the well-known DEHP algebra relation \cite{DEHP,sandow}:
\begin{equation} \label{eq:telre}
 DE-qED=(1-q)(D+E).
\end{equation}
On the boundaries, the relations become
\begin{eqnarray}
\llangle W|\left[B \left(\begin{array}{c}E\\D
 \end{array}\right)-(q-1)\left(\begin{array}{c}1\\-1
 \end{array}\right)\right]=0, \quad \left[\overline{B} \left(\begin{array}{c}E\\D
 \end{array}\right)+(q-1)\left(\begin{array}{c}1\\-1
 \end{array}\right)\right] |V\rrangle=0\;,
\end{eqnarray}
which are equivalent to the boundary relations \cite{DEHP,sandow}
\begin{equation}\label{eq:Bou}
\llangle W|\left(\alpha E-\gamma D+q-1 \right)=0, \quad \left(\delta E-\beta D+1-q \right) |V\rrangle=0.
\end{equation}
It remains to check the existence of non zero operators $E$ and $D$ and non zero vectors $\llangle W|$ and $|V\rrangle$ with $\llangle W|V\rrangle\neq 0$ 
which satisfy \eqref{eq:telre} and \eqref{eq:Bou}.
One way to prove this existence is to construct an explicit representation of $E$ and $D$ as well as $\llangle W|$ and $|V\rrangle$ (see for example \cite{DEHP}).

\paragraph{TASEP model.}
One can take the limit $q\to0$, $\gamma=0$, $\delta=0$ in the relations \eqref{eq:tel} and \eqref{eq:Bou} to obtain the relations defining the algebra and boundary conditions for TASEP.
These relations are equivalent to the well known relations \cite{DEHP}
\begin{equation}
 DE=D+E
\end{equation}
in the bulk, and for the boundaries
\begin{equation}
 \llangle W|\left(\alpha E-1 \right)=0, \quad \left(\beta D-1 \right) |V\rrangle=0.
\end{equation}

\paragraph{SSEP model.}
The limit $q\to1$ is not as straightforward as for TASEP. One has to modify the relations \eqref{eq:tel} and \eqref{eq:Bou}:
\begin{eqnarray}
 w^{SSEP}  \left(\begin{array}{c}E\\D
 \end{array}\right) \otimes \left(\begin{array}{c}E\\D
 \end{array}\right) = \left[ \left(\begin{array}{c}E\\D
 \end{array}\right) \otimes \left(\begin{array}{c}1\\-1
 \end{array}\right)-\left(\begin{array}{c}1\\-1
 \end{array}\right) \otimes  \left(\begin{array}{c}E\\D
 \end{array}\right)\right].
\end{eqnarray}
which is equivalent to
\begin{equation}
 ED-DE=D+E\ .
\end{equation}
For the boundaries, one gets
\begin{eqnarray}
\llangle W|\left[B \left(\begin{array}{c}E\\D
 \end{array}\right)-\left(\begin{array}{c}1\\-1
 \end{array}\right)\right]=0, \quad \left[\overline{B} \left(\begin{array}{c}E\\D
 \end{array}\right)+\left(\begin{array}{c}1\\-1
 \end{array}\right)\right] |V\rrangle=0\ ,
\end{eqnarray}
which are equivalent to
\begin{equation}
\llangle W|\left(\alpha E-\gamma D+1 \right)=0, \quad \left(\delta E-\beta D-1 \right) |V\rrangle=0.
\end{equation}

\section{Integrability: Yang-Baxter equation and reflection equation \label{sec:integrability}}

In this section, we review the main ingredients, the $R$-matrix and the $K$-matrix, that are needed to prove the integrability of  models with open boundaries.
These notions are well-known in the context of  quantum integrable models but we present them in the language of out-of-equilibrium statistical physics.
We provide also the explicit forms of the R-matrices and the $K$-matrices in the cases of the ASEP, SSEP and TASEP models.

\subsection{Yang-Baxter equation and $R$-matrix\label{sec:Rmat}}

The main object to deal with the bulk of an integrable model is called the $R$-matrix: it is a solution of the well-known  
Yang-Baxter equation \cite{yang,baxter} (written in $\CC^2 \otimes\CC^2 \otimes\CC^2$)
\begin{eqnarray}
 \label{eq:ybe}
 R_{12}(\frac{x_1}{x_2})\ R_{13}(\frac{x_1}{x_3})\ R_{23}(\frac{x_2}{x_3})\ =\ 
 R_{23}(\frac{x_2}{x_3})\ R_{13}(\frac{x_1}{x_3})\ R_{12}(\frac{x_1}{x_2})\ ,
\end{eqnarray}
where the subscripts indicate in which space the $R$-matrix acts non trivially:
$R_{12}=R\otimes 1$, $R_{23}=1\otimes R$,... (see section \ref{sec:notation}). The $R$-matrix $R(x)$ is a function of a parameter $x$ called the spectral parameter.

\begin{figure}[htbp]
\vspace{-3.5mm}
\begin{center}
\includegraphics[width=120mm]{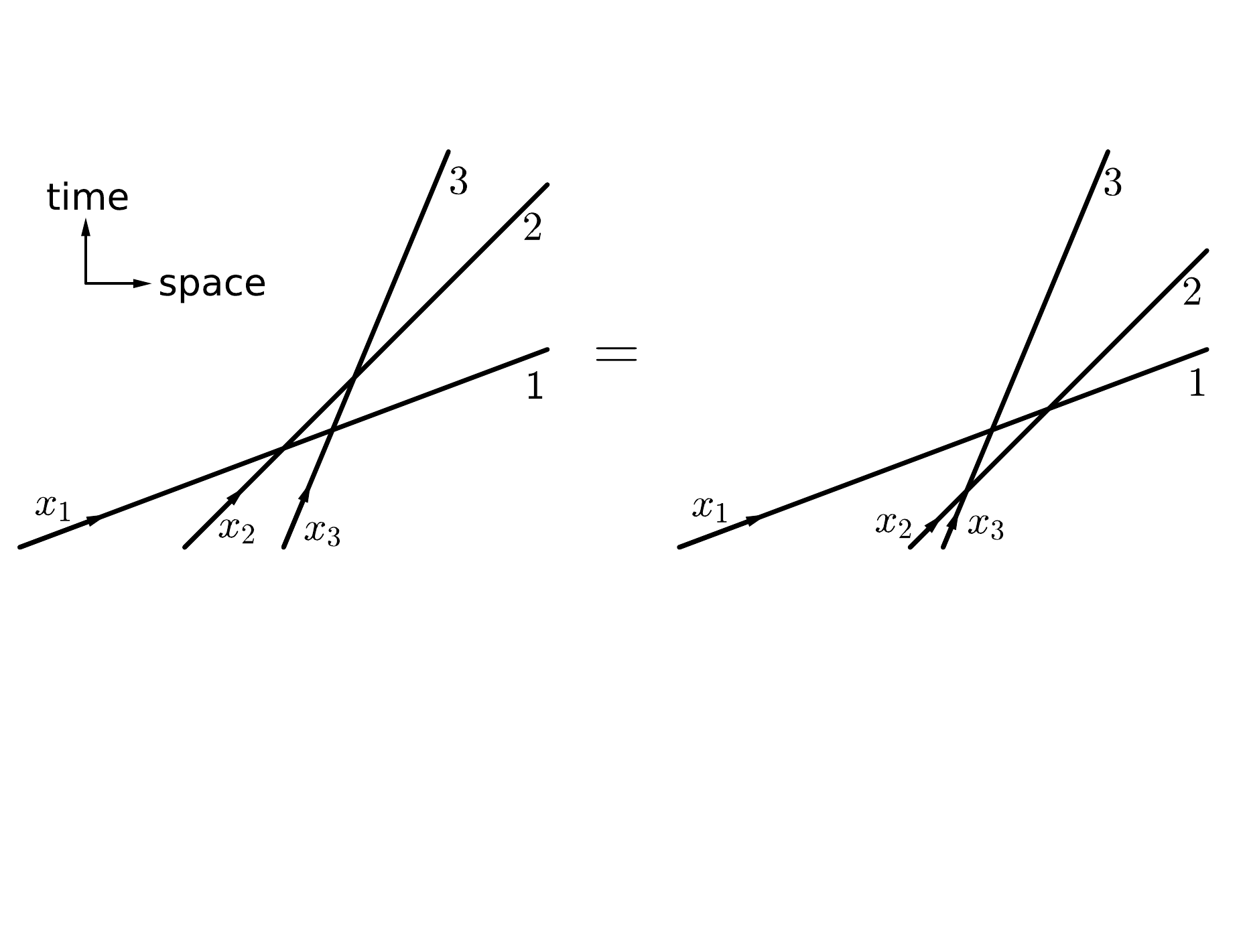}
\vspace{-35mm}
\caption{Schematic representation of the Yang-Baxter equation: the 3 particle scattering matrix is built out of 
2 particle scattering matrices and should not depend on the initial position of the particles (only on their relative order). 
The product of the 2 particle scattering matrices is ordered by time. \label{fig:ybe}}
\end{center}
\end{figure}

The $R$-matrix allows one to construct the transfer matrix for the model with periodic boundary conditions. We do not recall here 
this construction for the periodic case but, in section \ref{transfer_matrix}, we construct the transfer matrix for the open case.
The Yang-Baxter equation has a nice pictorial representation (see figure \ref{fig:ybe}) coming from integrable quantum field theory: 
the $R$-matrix $R(x_1/x_2)$ is the diffusion matrix between two particles with rapidities $x_1$ and $x_2$, the integrability in this context 
is the fact that the diffusion matrix of 3 particles factorizes 
in this $R$-matrix and the Yang-Baxter equation is the consistency relation of this factorization.

Usually, one requires also that the $R$-matrix satisfies supplementary conditions: 
\begin{itemize}
\item Unitarity property
\begin{eqnarray}
\label{eq:unitary}
 R_{12}(x)\,R_{21}(\frac1x)=1
\end{eqnarray}
\item Regularity property
\begin{eqnarray}
\label{eq:regularity}
 R(1)=P, 
\end{eqnarray}
where $P$ is the permutation operator. This property is important since it allows one to get local Markov matrix when 
one takes the derivative of the transfer matrix (which will be defined in subsection \ref{transfer_matrix}). Remark that once
the Yang-Baxter equation is ensured, unitarity is implied by the regularity relation.
\item Crossing unitarity property 
\begin{equation}
\label{eq:crossing}
 R_{12}(x)^{t_2}\,U_2\,R_{21}(\frac1{xQ})^{t_2}\,U_2^{-1} = \lambda(x), 
\end{equation}
where ${}^{t_2}$ is the transposition in the second component of the tensor space. This is a key relation to show
that the transfer matrices introduced in \cite{sklyanin} commute. 
 However this property is not always satisfied: the $R$-matrix associated to 
TASEP model does not satisfy the crossing unitarity property. 
\item Local jump operator property
\begin{equation}\label{eq:jump}
 P\,R'(1)= \rho\ w
\end{equation}
where $.'$ stands for the derivative with respect to the spectral parameter, $\rho$ is a constant 
and $w$ is the local jump operator appearing in the Markov matrix \eqref{eq:W}.

\item Markovian property: there exists a vector $v(x)$ such that
\begin{eqnarray}\label{eq:marmr}
 R_{12}(\frac{x_1}{x_2})\ v_1(x_1)\ v_2(x_2)\ =\ v_1(x_1)\ v_2(x_2),
\end{eqnarray}
and
\begin{equation}\label{eq:marml}
 \langle 1| \otimes \langle 1| R(x)=\langle 1| \otimes \langle 1|,
 \mb{with}  \langle 1|=(1\,,\,1).
\end{equation}
This relation is unusual in the context of integrable systems but it seems natural to add it when studying Markovian 
process: it is a sufficient condition for the Markovian matrix obtained from the $R$-matrix 
to have a vanishing eigenvalue. 
Indeed, deriving relation \eqref{eq:marml} with respect to $x$ and taking $x=1$, we get
\begin{equation}
 \langle 1| \otimes \langle 1| w=0\;.
\end{equation}
Therefore, the probability is conserved. To get a Markov matrix, one must still check that the off-diagonal entries are positive.  
We will show in section \ref{sec:ZF} that this property allows us to relate quantum groups and ZF algebras. As a consequence, properties already 
proved for the quantum group are automatically satisfied for the ZF algebra. 
This Markovian property replaces the property
$R^\dagger\ R=1$ appearing in the context of quantum field theory.
However, it is not essential for our purpose and most of our calculations can be performed without assuming this property.
\end{itemize}

To end this subsection, let us emphasize that solving the Yang-Baxter equation is a difficult task since it is a cubic equation in the entries of the $R$-matrix. 
The first examples have been obtained by C.N.Yang \cite{yang} and R.J.Baxter \cite{baxter}. A 
brute-force computation to get the solutions is still an open problem and is an active domain of research 
(see e.g. \cite{KulS,martins,MP,khac,FFR}). 
In this context, the use of the quantum group can be very fruitful since it
gives an algebraic framework to the Yang-Baxter equation.
Instead of solving the Yang-Baxter equation, 
one can use the intertwiner relation which 
is linear in $R$ to get the solution of the Yang-Baxter equation \cite{jimbo}
or one can represent the universal Yang-Baxter equation to get a new solution for each representation of the quantum group
\cite{drinfeld}. Nevertheless, to use these methods, one must know and study the underlying quantum groups.

\subsection{Reflection equation}

\begin{figure}[htbp]
\begin{center}
\includegraphics[width=120mm]{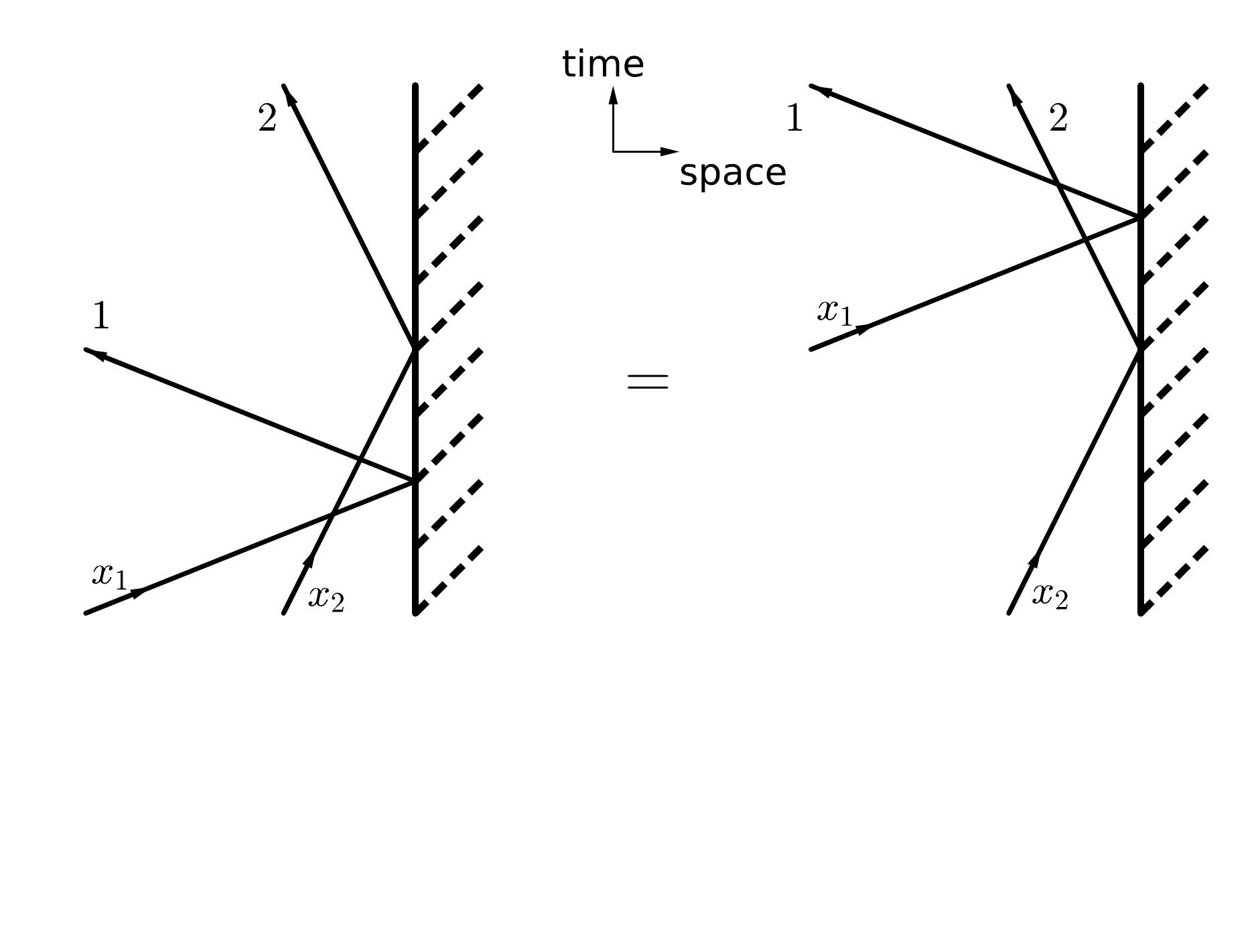}
\vspace{-35mm}
\end{center}
\caption{Schematic representation of the reflection equation. \label{fig:re}}
\end{figure}

To deal with boundary conditions which preserve the integrability of the model, a supplementary equation, called reflection equation is needed \cite{Cher,sklyanin}. 
It is a compatibility equation between the bulk, encoded in the $R$-matrix, and the boundary, encoded in the so-called boundary matrices or $K$-matrices.
The reflection equation  is written as follows
\begin{equation}
\label{eq:re}
 R_{12}(\frac{x_1}{x_2})\, K_1(x_1)\, R_{21}(x_1 x_2)\, K_2(x_2)=
 K_2(x_2)\,R_{12}(x_1 x_2)\,K_1(x_1)\,R_{21}(\frac{x_1}{x_2})\;.
\end{equation}
As for the Yang-Baxter equation, there exists a nice graphical interpretation for the reflection equation
coming from  quantum field theory (see figure \ref{fig:re}): the $K$-matrix $K(x)$ is the scattering matrix of a particle with rapidity $x$ on the boundary
and the reflection equation is the compatibility for the factorization of the scattering of two particles on the boundary. 

Here we will only be interested in the solutions satisfying
\begin{itemize}
\item Unitarity property
\begin{equation}
\label{eq:unitary_K}
 K(x)\,K(\frac1x)=1.
\end{equation}
\item Regularity property
\begin{equation}
\label{eq:regularity_K}
 K(1)=1.
\end{equation}
\item Boundary jump operator property
\begin{equation} \label{eq:jump_K}
 K'(1)= 2\rho\,\epsilon\,B,
\end{equation}
where $B$ is the boundary jump operator, $\rho$ is the same coefficient as the one appearing in \eqref{eq:jump}
and $\epsilon=\pm 1$. For the $K$-matrix associated to the left (respectively right) boundary, $B$ is the left (respectively the right)
boundary jump operator and $\epsilon=1$ (respectively $\epsilon=-1$).
\item Markovian property: there exists a vector $u(x)$ and $\langle 1|=(1\,,\,1)$ such that
\begin{equation}
 K(x)\,u(\frac1x)=u(x),
\mb{and}
 \langle 1|K(x)=\langle 1|.
\end{equation}
\end{itemize}

As for the Yang-Baxter equation, the first task consists in finding the solution of the reflection equation for a given $R$-matrix and, when possible, 
to give a classification of these solutions.
The first solutions to the reflection equation have been found in \cite{Cher,sklyanin}. Then, a classification has been done in \cite{vega}
for the $R$-matrix associated to the SSEP model and in \cite{vega,GZ} for the ASEP model (see section \ref{sec:exint} for their explicit form).
For more complicated R-matrices, there exist also some solutions but the complete classification is still an open problem and the subject of many
researches \cite{AR,lima,gand,AACDFR1,AACDFR2}.

Even though we do not give detail here, we would like also to mention that the reflection equation is very useful in the context of the quantum group.
Indeed, it allows one to obtain subalgebras (in fact Hopf coideals) of quantum groups, see e.g. \cite{Olsh,MNO,MRS,MR,Mu}. When the algebraic structure of these coideals is known, the use an 
intertwiner relation (linear in $K$) provides solutions of the reflection equation \cite{DM,DMS,BC}.

\subsection{Transfer matrix \label{transfer_matrix}}

In this section, we recall why the reflection equation is sufficient to get boundaries which preserve the integrability
and how to construct the corresponding transfer matrix \cite{sklyanin}.
Indeed, in order to prove the integrability of the Markov matrix $M$, one must construct a more general operator, called the transfer matrix $t(x)$, that
depends on a spectral parameter $x$. 
The transfer matrix $t(x)$ commutes for different values of the spectral parameter
(\textit{i.e.} $[t(x),t(x')]=0$) and its first derivative is proportional to $M$ (\textit{i.e.} $t'(1)\propto M$). 
Hence the transfer matrix is a generating function for the conserved charges of the model.

\begin{figure}[htbp]
\begin{center}
\vspace{-0.4cm}
\includegraphics[width=120mm]{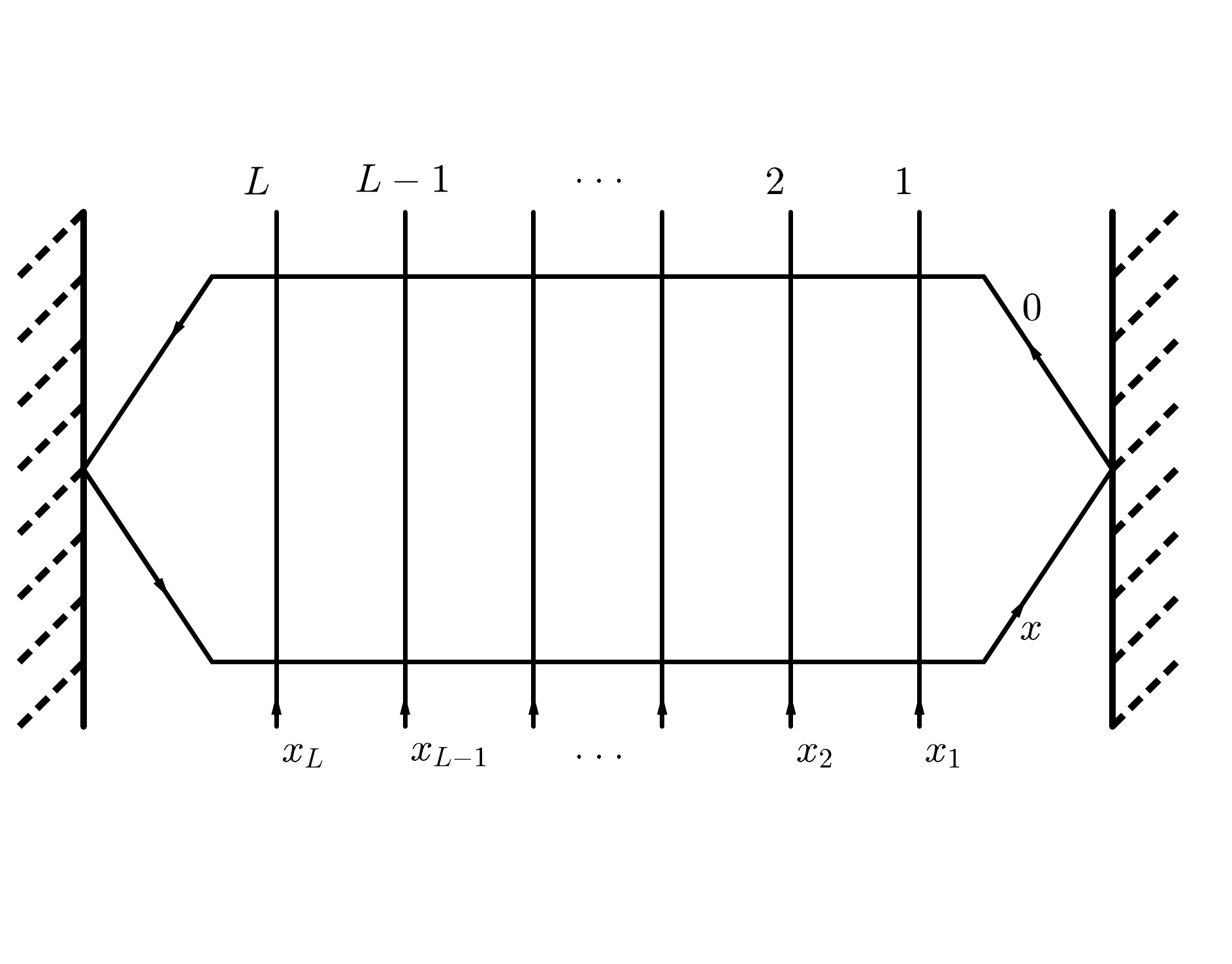}
\vspace{-1.7cm}
\caption{The transfer matrix in the open case \label{fig:t}}
\end{center}
\end{figure}

For the case with boundaries, the transfer matrix is given in terms of the $R$-matrix and two $K$-matrices (one for each boundary) \cite{sklyanin}:
\begin{equation}\label{eq:tb}
 t(x)=\frac{1}{tr(\widetilde K(1))} \ tr_0\Big( \widetilde K_0(x)\ R_{0L}(x)\dots R_{01}(x)\ K_0(x)\ R_{10}(x)\dots R_{L0}(x)  \Big)\;,
\end{equation}
where $K(x)$ is a solution of the reflection equation \eqref{eq:re} and $\widetilde{K}(x)$ is a solution of the so-called dual reflection equation 
\begin{equation} \label{eq:re_dual}
 \widetilde{K}_2(x_2)\,\left(R_{21}^{t_1}(x_1x_2)^{-1}\right)^{t_1}\,\widetilde{K}_1(x_1)\,R_{21}(\frac{x_2}{x_1})
 =R_{12}(\frac{x_2}{x_1})\,\widetilde{K}_1(x_1)\,\left(R_{12}^{t_2}(x_1x_2)^{-1}\right)^{t_2}\,\widetilde{K}_2(x_2).
\end{equation}
Using the crossing unitarity property \eqref{eq:crossing} it can be rewritten as follows\\
\begin{equation}
 \widetilde{K}_2(x_2)\,U_1^{-1}\,R_{12}(\frac1{x_1x_2Q})\,U_1\,\widetilde{K}_1(x_1)\,R_{21}(\frac{x_2}{x_1})
 =R_{12}(\frac{x_2}{x_1})\,\widetilde{K}_1(x_1)\,U_2^{-1}\,R_{21}(\frac1{x_1x_2Q})\,U_2\,\widetilde{K}_2(x_2).
\end{equation}
Let us mention that there exist several bijective maps between the solutions of the reflection equation and the solutions of the dual reflection equation.
In the following we will often make use of the following one. If $\bar K$ is a solution to the reflection equation
\eqref{eq:re} associated to the $R$-matrix $\widetilde{R}_{12}(x)=R_{21}(1/x)=R_{12}(x)^{-1}$,
\begin{equation} \label{eq:re_bis}
 R_{21}(\frac{x_2}{x_1})\bar K_1(x_1)R_{12}(\frac1{x_1x_2})\bar K_2(x_2)=
 \bar K_2(x_2)R_{21}(\frac1{x_1x_2})\bar K_1(x_1)R_{12}(\frac{x_2}{x_1}),
\end{equation}
then the matrix $\widetilde{K}(x)$ defined by
\bea \label{eq:Ktilde}
 \widetilde K_1(x) \, =\, tr_0\Big(\bar{K}_0(\frac1x)\big((R_{01}(x^2)^{t_1})^{-1}\big)^{t_1}P_{01} \Big) 
\, =\, \frac{1}{\lambda(x^2)}\ tr_0\Big( \bar{K}_0(\frac1x)U_1^{-1}R_{10}(\frac1{x^2Q})U_1P_{01} \Big),\quad
\eea
is a solution to the dual reflection equation \eqref{eq:re_dual} \cite{sklyanin}\footnote{In \cite{sklyanin},
this property was shown assuming that $R_{12}=R_{21}$, but the generalization is straightforward.}.
The second equality in \eqref{eq:Ktilde} is obtained using the crossing unitarity property \eqref{eq:crossing}.
Remark that $\widetilde{K}$ (and the dual reflection equation) can be defined only when $R_{01}^{t_1}$ is invertible.
This will not be true for the TASEP model, see section \ref{sec:et}. 
One can invert the previous relation and get
\begin{equation} \label{eq:Kbar}
 \bar K_1(x)= tr_0\Big( \widetilde{K}_0(\frac1x)R_{01}(\frac1{x^2})P_{01} \Big)\;.
\end{equation}
The transfer matrix \eqref{eq:tb} has a nice graphical interpretation (see figure \ref{fig:t}): a test particle with rapidity $x$ runs in the 
lattice from site $L$ to site $1$ (scattering with each site), reflects on the boundary closed to site $1$, comes back from site $1$ to $L$ 
and reflects on the boundary closed to $L$.

As mentioned previously, we can get a Markov matrix from this transfer matrix 
\begin{equation}
\frac{1}{2\rho} \left. \frac{dt(x)}{dx}\right|_{x=1}=\frac{1}{2\rho} K_1'(1)+\frac{1}{\rho}\sum_{k=1}^{L-1}P_{k,k+1}R'_{k,k+1}(1)
 -\frac{1}{2\rho}\bar K_L'(1)=M.
\end{equation}
It was also shown in \cite{sklyanin} that if $K$ satisfies the reflection equation \eqref{eq:re} and $\widetilde{K}$ satisfies the 
dual reflection equation \eqref{eq:re_dual},
then the transfer matrix commutes for different values of the spectral parameter
\begin{equation}
 [t(x),t(x')]=0.
\end{equation}
Then the transfer matrix is the generating function of operators commuting with $M$ which proves the 
integrability\footnote{One must also 
show that the number of independent conserved quantities is equal to the number of degree of freedom. This problem 
can be tackled only case by case.} of $M$. Let us emphasize that the crossing unitarity of the $R$-matrix is crucial in \cite{sklyanin} to prove
the commutativity of the transfer matrix. We come back to this point in section \ref{sec:et}.

Then, a natural question arises: whether the stationary state $\steady$ of the Markov matrix given by the matrix ansatz \eqref{eq:O} is also an eigenvector for the
transfer matrix $t(x)$ and what is the associated eigenvalue. We answer this question in section \ref{sec:ti} for the different models we have presented.

\subsection{Examples \label{sec:exint}}

In this section, we provide the explicit forms of the R-matrices associated to the ASEP, TASEP and SSEP models.
We give also the $K$-matrices which leads to  integrable markovian boundaries. 

\subsubsection{ASEP model \label{sec:eax}}

The $R$-matrix associated to the ASEP model is given by
\begin{equation}\label{eq:RASEP}
R(x)=\left( \begin {array}{cccc} 
1&0&0&0\\ 
0&\frac{(x-1)q}{qx-1}&\frac{(q-1)x}{qx-1}&0\\
0&\frac{q-1}{qx-1}&\frac{x-1}{qx-1}&0\\
0&0&0&1
\end {array} \right) .
\end{equation}
It satisfies the Yang-Baxter equation, the unitarity, the regularity, the local jump properties (with $\rho=1/(q-1)$ and $w$ given in \eqref{eq:localwASEP}) 
and the crossing unitarity property with  
\begin{equation}
Q=q^2\,, \quad \lambda(x)=\frac{(x-1)(q^2x-1)}{(qx-1)^2}\quad \mbox{and}\quad 
U=\left( \begin {array}{cc} 1&0\\  0&q\end {array} \right).
\end{equation}
The Markovian property is verified for the vector
\begin{equation}
 v(x)=f(x)\left( \begin {array}{c} 
ax\\ 
b
\end {array} \right),
\end{equation}
where $a$, $b$ are arbitrary constants and $f$ is an arbitrary function. 

We introduce a solution $K(x)$ of the reflection equation \eqref{eq:re} 
\begin{equation}\label{eq:Kae}
K(x)=\left( \begin {array}{cc} 
\displaystyle{\frac{(-x \alpha+x \gamma+q+\alpha-\gamma-1)x}{x^2 \gamma+q x+x \alpha-x \gamma-x-\alpha}}&\displaystyle{\frac{(x^2-1)\gamma}{x^2\gamma+qx+x\alpha-x\gamma-x-\alpha}}\\[2ex]
\displaystyle{\frac{\alpha(x^2-1)}{x^2\gamma+qx+x\alpha-x\gamma-x-\alpha}}&\displaystyle{-\frac{(-qx-x\alpha+x\gamma+x+\alpha-\gamma)}{x^2\gamma+qx+x\alpha-x\gamma-x-\alpha}}
\end {array} \right),
\end{equation}
and a solution $\bar K(x)$ of the reflection equation \eqref{eq:re_bis}
\begin{equation}\label{eq:Kab}
\bar K(x)= \left( \begin {array}{cc} 
\displaystyle{\frac{(x \delta-x \beta+q-\delta+\beta-1)x}{-x^2 \beta+qx-x \delta+x \beta-x+\delta}}&\displaystyle{-\frac{(x^2-1)\beta}{-x^2\beta+qx-x\delta+x\beta-x+\delta}}\\[2ex]
\displaystyle{-\frac{(x^2-1)\delta}{-x^2\beta+qx-x\delta+x\beta-x+\delta}}&\displaystyle{\frac{qx-x\delta+x\beta-x+\delta-\beta}{-x^2\beta+qx-x\delta+x\beta-x+\delta}}
\end {array} \right).
\end{equation}
These matrices satisfy the unitarity and regularity relations. As explained previously, 
they encode the boundaries \eqref{eq:localwASEP} through their derivatives (boundary jump operator property)
\begin{equation}
 \frac{1}{2\rho}K'(1)=B, \quad -\frac{1}{2\rho}\bar K'(1)=\overline{B},
\end{equation}
with $\rho=1/(q-1)$.
The matrix $\widetilde K(x)$ corresponding to the matrix $\bar K(x)$ given in \eqref{eq:Kab} is
 \begin{equation} \label{Ktilde_asep}
  \widetilde K(x)= \frac{qx^2-1}{(\delta x +\beta)(x-1)+(q-1)x}\left( \begin {array}{cc} 
\frac{qx(q-1-\delta+\beta)+\delta-\beta}{q^2x^2-1}&\beta\\[2ex]
\displaystyle\frac{\delta}{q}&\frac{x(qx(\delta-\beta)+q-1-\delta+\beta)}{q^2x^2-1}
\end {array} \right).
 \end{equation}

Let us mention that there exists a more general $K$-matrix given by \cite{vega} 
\begin{equation}\label{eq:Kaeg}
K(x)=\left( \begin {array}{cc} 
\displaystyle{\frac{(-x \alpha+x \gamma+q+\alpha-\gamma-1)x}{x^2 \gamma+q x+x \alpha-x \gamma-x-\alpha}}
&\displaystyle{e^{-s}\ \frac{(x^2-1)\gamma}{x^2\gamma+qx+x\alpha-x\gamma-x-\alpha}}\\[2ex]
\displaystyle{e^{s}\ \frac{\alpha(x^2-1)}{x^2\gamma+qx+x\alpha-x\gamma-x-\alpha}}&\displaystyle{-\frac{(-qx-x\alpha+x\gamma+x+\alpha-\gamma)}{x^2\gamma+qx+x\alpha-x\gamma-x-\alpha}}
\end {array} \right),
\end{equation}
which gives the boundary matrix
 \begin{eqnarray}\label{soluBg}
 B =\left( \begin {array}{cc} 
-\alpha&\gamma e^{-s}\\ 
 \alpha e^{s}&-\gamma
\end {array} \right)
\;.
\end{eqnarray}
This general matrix does not satisfy the Markovian property. However, it is very useful in the context of  out-equilibrium statistical physics since 
the eigenvalues of this matrix provide the full distribution of the total current and its large deviation behavior.
Although the model stays integrable with this general boundary, the computation of the spectrum and of these eigenvectors is still an active domain of research.
We discuss this point in detail in section \ref{sec:ba}.

\subsubsection{TASEP model \label{sec:et}}

The $R$-matrix associated to the TASEP model is obtained by taking the limit $q=0$ in the previous expression:
\begin{equation}
R(x)=\left( \begin {array}{cccc} 
1&0&0&0\\ 
0&0&x&0\\
0&1&1-x&0\\
0&0&0&1
\end {array} \right) .
\end{equation}
It satisfies the Yang-Baxter equation, the unitarity, the regularity, local jump operator property (with $\rho=-1$ and $w$ given by \eqref{eq:wta}) and the Markovian property for the vector
\begin{equation}
v(x)=f(x)\left( \begin {array}{c} 
ax\\ 
b
\end {array} \right),
\end{equation}
where $a$, $b$ are arbitrary constants and $f$ is an arbitrary function. 
However $R_{12}(x)^{t_2}$ is not invertible anymore so it does not satisfy the crossing unitarity property. 

One can take also the limit $q=0,\gamma=0,\delta=0$ in the expressions \eqref{eq:Kae}, \eqref{eq:Kab} to get the reflection matrices for TASEP
\begin{equation} \label{K_tasep}
K(x)=\left( \begin {array}{cc} 
\displaystyle{\frac{(-x \alpha+\alpha-1)x}{x \alpha-x-\alpha}}&0\\[2ex]
\displaystyle{\frac{\alpha(x^2-1)}{x\alpha-x-\alpha}}&1
\end {array} \right)\quad
\text{and}\qquad
\bar K(x)= \left( \begin {array}{cc} 
1&\displaystyle{-\frac{(x^2-1)\beta}{-x^2\beta+x\beta-x}}\\[2ex]
0&\displaystyle{\frac{x\beta-x-\beta}{-x^2\beta+x\beta-x}}
\end {array} \right).
\end{equation}
The unitarity and the regularity properties still hold for $K$ and $\bar K$ and one has the boundary jump operator property \eqref{eq:jump_K}
with $\rho=-1$ and the left/right boundary jump operator given by \eqref{eq:wta}.

We would like to take the limit $q\to0$ of the ASEP case to construct the transfer matrix for TASEP. One has to be very careful because for $q=0$, the formula \eqref{eq:Kbar} is not invertible anymore and the dual reflection equation \eqref{eq:re_dual}
is not defined. However, one can check that the expression \eqref{Ktilde_asep} of $\widetilde{K}$  is regular
and well-defined if we set $\delta=0$ before taking the limit $q\to0$. It gives
\begin{equation} 
  \widetilde K(x)= \frac{1}{x(\beta-1)-\beta}\left( \begin {array}{cc} 
-\beta&-\beta\\ 
0&x(\beta-1)
\end {array} \right).
 \end{equation}
Using this $\widetilde{K}$ matrix and the $K$ matrix in \eqref{K_tasep}, we can construct the transfer matrix \eqref{eq:tb}. The commutation of
this transfer matrix for two different spectral parameters is ensured by the fact that the transfer matrix of the ASEP model for $\delta=0$ and $\gamma=0$ goes
to these of TASEP model as $q$ goes to $0$. This is to the best of our knowledge the first proof of the integrability (in the sense that there exists
a family of commuting transfer matrices) of the open TASEP model since it does not fit directly in the theory developed in \cite{sklyanin}.
Let us mention \cite{DFIL,AM} where there are attempts to study integrable models with boundaries based on a $R$-matrix without crossing unitarity. However, 
they are case-by-case methods and, as far as we know, there is no general method.

\subsubsection{SSEP model}

The SSEP $R$-matrix is given by
\begin{equation}\label{eq:RSSEP}
R(x)=\frac{x+P}{x+1}=\left( \begin {array}{cccc} 
1&0&0&0\\ 
0&\frac{x}{x+1}&\frac{1}{x+1}&0\\
0&\frac{1}{x+1}&\frac{x}{x+1}&0\\
0&0&0&1
\end {array} \right)\ .
\end{equation}
It can be obtained from the ASEP $R$-matrix through a scaling limit $q \rightarrow 1$. Let us set $q=e^{\lambda}$ and $z=e^{x\lambda}$. Then one has the following result
\begin{equation}
 \lim_{\lambda \to 0} R^{ASEP}(z)=R^{SSEP}(x).
\end{equation}
Let us remark that in this limit, we switch from a multiplicative spectral parameter $z$ (in ASEP) to an additive one $x$ (in SSEP). \\

The SSEP $R$-matrix satisfies the Yang-Baxter equation, the unitarity, regularity, local jump operator, the crossing unitarity and  Markovian properties but with additive arguments:
\begin{eqnarray}
 \label{eq:ybeadd}
 &&R_{12}(x_1-x_2)\ R_{13}(x_1-x_3)\ R_{23}(x_2-x_3)\ =\ R_{23}(x_2-x_3)\ R_{13}(x_1-x_3)\ R_{12}(x_1-x_2)\qquad\ \\
 &&R_{12}(x)R_{21}(-x)=1\,,\quad R(0)=P\,,\quad R_{12}(x)^{t_2}R_{21}(-x-2)^{t_2}= x(x+2)/(x+1)^2\\
 &&P\,R'(0)= w\,,\quad R_{12}(x_1-x_2)\ v_1(x_1)\ v_2(x_2)\ =\ v_1(x_1)\ v_2(x_2),
\end{eqnarray}
where $w$ is given by \eqref{eq:ws} and
\begin{equation}
 v(x)=f(x)\left( \begin {array}{c} a\\ b\end {array} \right),
\end{equation}
with $a$, $b$ arbitrary constants and $f$ an arbitrary function.
One has the extra relation $R_{12}(x)=R_{21}(x)$.

For SSEP models, the reflection equation has also additive arguments
\begin{equation} \label{eq:readd}
 R_{12}(x_1-x_2) K_1(x_1) R_{21}(x_1+x_2) K_2(x_2)=K_2(x_2)R_{12}(x_1+x_2)K_1(x_1)R_{21}(x_1-x_2)\;.
\end{equation}
In the context of Markovian process, the solutions  are given by
\begin{equation}\label{eq:Kds}
K(x)= \frac{1}{x(\alpha+\gamma)+1}\left( \begin {array}{cc} 
x(\gamma-\alpha)+1&2x\gamma\\ 
2x\alpha&x(\alpha-\gamma)+1
\end {array} \right)
\end{equation}
 for equation \eqref{eq:readd} and
\begin{equation}\label{eq:bKSSEP}
\bar K(x)= \frac{1}{x(\delta+\beta)-1}\left( \begin {array}{cc} 
x(\beta-\delta)-1&2x\beta\\ 
2x\delta&x(\delta-\beta)-1
\end {array} \right)
\end{equation}
for equation \eqref{eq:re_bis} with additive arguments.
As in the previous cases, $K$ and $\bar K$ satisfy the unitarity property $K(x)K(-x)=1$, the regularity property $K(0)=1$
and are related to the boundaries through their derivatives (boundary jump operator property)
\begin{equation}
 \frac{1}{2}K'(0)=B, \quad -\frac{1}{2}\bar K'(0)=\overline{B}.
\end{equation}
 The matrix $\widetilde K(x)$ corresponding to the matrix $\bar K(x)$ given in \eqref{eq:bKSSEP} is
 \begin{equation}\label{eq:Kts}
  \widetilde K(x)= \frac{2x+1}{x(\delta+\beta)+1}\left( \begin {array}{cc} 
\frac{(x+1)(\beta-\delta)+1}{2(x+1)}&\beta\\ 
\delta&\frac{(x+1)(\delta-\beta)+1}{2(x+1)}
\end {array} \right).
 \end{equation}

\section{Quantum group \label{quantum_group}}

\subsection{FRT formalism for quantum group\label{sec:frt}}

There exist various presentations of quantum groups. Here, we use the one
given in \cite{FRT} and so-called FRT presentation, in reference to the authors of \cite{FRT} 
(or RTT presentation, named from the form of relation \eqref{rttg}). It is based on the $R$-matrix and is suitable
for the link with the matrix ansatz we present in section \ref{sec:link}. We provide in this section the general definitions and gives the examples related to ASEP 
models in the next subsections.

The general idea consists in giving the generators of the considered algebra as 
formal series\footnote{The precise development of 
these series depends on the considered quantum groups, see the examples.} 
${\cal A}(x)$, ${\cal B}(x)$, ${\cal C}(x)$ and ${\cal D}(x)$ gathered in a two by two matrix
\begin{equation}
 T(x)=\left(\begin{array}{c c}
    {\cal A}(x)&{\cal B}(x)\\
    {\cal C}(x)& {\cal D}(x)
   \end{array}\right)\;.
\end{equation}
Then, the commutation relations are encoded in the following relation 
\begin{equation}\label{rttg}
R_{12}(x_1/x_2)\,T_1(x_1)\,T_2(x_2) = T_2(x_2)\,T_1(x_1)\,R_{12}(x_1/x_2)
\end{equation}
where $R$ is a $R$-matrix and the subscript on the T's indicates in which space the two by two matrix acts. 
The $R$-matrix may be seen as the constant structures for the algebra and the Yang-Baxter equation is sufficient
to guarantee the associativity for the algebra. 

We would like to mention that there exists some realisation of this infinite dimensional algebra by a finite dimensional one, 
when we replace the series ${\cal A}(x)$,... by some finite sum (i.e. we set an infinite number of generators to zero).
This realisation is called an evaluation map (see section \ref{sec:eqg} for examples).
Let us remark that this operation can lead to a trivial representation since the commutation relations and the fact that some generators are put to 0 
may imply that the other generators vanish too. Then, when we say that there exists an evaluation map, 
it is understood implicitly that some of the generators are not zero.

Let us also mention the simple representation 
given by the so-called monodromy matrix
\begin{equation}
 T_0(x)=R_{0L}(x)\dots R_{02}(x) R_{01}(x)\;.
\end{equation}

The RTT presentation is important in the context of integrable systems since it gives the commutation relations between the operators 
needed for the algebraic Bethe ansatz. This method allows one to compute 
the eigenvalues and the eigenvectors for periodic integrable systems \cite{QISM} (see \cite{how} for a review).\\

From this algebra, we can construct a subalgebra, called reflection algebra and based on the reflection equation.
Indeed, let us define
\begin{equation}
 B(x)\ =\ T(x)\ K(x)\ T(1/x)^{-1},
\end{equation}
where $K(x)$ is a solution of the reflection equation \eqref{eq:re}. Then, one gets
\begin{equation}
\label{eq:rea}
 R_{12}(x_1/x_2) B_1(x_1) R_{21}(x_1 x_2) B_2(x_2)=B_2(x_2)R_{12}(x_1 x_2)B_1(x_1)R_{21}(x_1/x_2)\;.
\end{equation}
The reflection algebra is the cornerstone to generalize the algebraic Bethe ansatz to open integrable models \cite{sklyanin}. 
Up to recently, this method worked only for diagonal boundary matrices and
could not be applied for Markov matrix with open boundaries. However, in \cite{BCR,PL}, the algebraic Bethe ansatz for triangular boundaries has been developed
and in \cite{BC13}, it has been developed for the SSEP model (\textit{i.e.} XXX model) with general boundaries. 
A direct comparison with the matrix ansatz solution would be very interesting.

\subsection{Zamolodchikov algebra\label{sec:ZF}}

Another type of subalgebra of the previous quantum groups is interesting for us. The generators of this subalgebra ${\cal X}_1(x)$ and ${\cal X}_2(x)$ are gather in a vector $A(x)$ 
and are linear combinations of the previous generators 
${\cal A}(x)$, ${\cal B}(x)$, ${\cal C}(x)$ and ${\cal D}(x)$. They read
\begin{equation}\label{eq:T-A}
 A(x)=\left(\begin{array}{c}{\cal X}_1(x)\\ {\cal X}_2(x)\end{array}\right)= T(x) v(x)\;,
\end{equation}
where $v(x)$ is the scalar vector appearing in the Markovian property \eqref{eq:marmr}. 
Explicitly if we write  $v(x)=\left(\begin{array}{c}v_1(x)\\ v_2(x)\end{array}\right)$, one gets
\begin{equation}
 {\cal X}_1(x)=v_1(x) {\cal A}(x)+v_2(x) {\cal B}(x)\quad\text{and}\qquad 
 {\cal X}_2(x)=v_1(x) {\cal C}(x)+v_2(x) {\cal D}(x)\;.
\end{equation}

The exchange relations for $A(x)$ are obtained easily from the RTT relation \eqref{rttg}. 
Indeed, if we act with $v_1(x_1)v_2(x_2)$
on the RTT relation and use the Markovian property, we get
\begin{equation}\label{eq:ZF}
 R_{12}(x_1/x_2) A_1(x_1) A_2(x_2)=A_2(x_2) A_1(x_1)\;.
\end{equation}
This relation introduced in \cite{ZF1} (and generalized in \cite{ZF2}) is well-known in the context of integrable quantum field theory
and is called the Zamolodchikov relation. It is a generalization of the commutation of bosons (obtained for $R(x)=1$) or the anticommutation 
of fermions (obtained for $R(x)=-1$). 

The associativity of the Zamolodchikov algebra is ensured by the Yang-Baxter equation. There are two ways of reversing the order of the product $A_3(x_3)A_2(x_2)A_1(x_1)$:
\begin{eqnarray}
 A_3(x_3)A_2(x_2)A_1(x_1)&=&R_{12}(\frac{x_1}{x_2})A_3(x_3)A_1(x_1)A_2(x_2)
 = R_{12}(\frac{x_1}{x_2})R_{13}(\frac{x_1}{x_3})A_1(x_1)A_3(x_3)A_2(x_2)\nonumber\\
 &=&R_{12}(\frac{x_1}{x_2})R_{13}(\frac{x_1}{x_3})R_{23}(\frac{x_2}{x_3})A_1(x_1)A_2(x_2)A_3(x_3),
\end{eqnarray}
or
\begin{eqnarray}
 A_3(x_3)A_2(x_2)A_1(x_1)&=&R_{23}(\frac{x_2}{x_3})A_2(x_2)A_3(x_3)A_1(x_1)
 = R_{23}(\frac{x_2}{x_3})R_{13}(\frac{x_1}{x_3})A_2(x_2)A_1(x_1)A_3(x_3)\nonumber\\
 &=&R_{23}(\frac{x_2}{x_3})R_{13}(\frac{x_1}{x_3})R_{12}(\frac{x_1}{x_2})A_1(x_1)A_2(x_2)A_3(x_3).
\end{eqnarray}

Another consistency relation is ensured by the unitarity property \eqref{eq:unitary} of the $R$-matrix: if one applies the Zamolodchikov relation twice
\begin{equation}
 A_2(x_2) A_1(x_1)=R_{12}(\frac{x_1}{x_2})\,A_1(x_1) A_2(x_2)=R_{12}(\frac{x_1}{x_2})\,R_{21}(\frac{x_2}{x_1})\,
 A_2(x_2) A_1(x_1).
\end{equation}

Relation \eqref{eq:T-A} expressing the Zamolodchikov algebra generators  
 as linear combinations of quantum group generators  is, to the best of our knowledge, a new result. The existence of this relation crucially 
 depends on the Markovian property \eqref{eq:marmr}. As already mentioned, this relation is nevertheless not essential for our purpose.
Other relations between Zamolodchikov algebra and quantum groups have been previously 
established\footnote{More precisely, in
these papers, they considered the Zamolodchikov-Faddeev algebra which is a generalization of the Zamolodchikov algebra.}: in \cite{ku,miki}   
the Zamolodchikov generators are quadratic expressions in the quantum group generators and in \cite{well-bred}, the quantum group generators are expressed as series in the Zamolodchikov ones.

Finally, let us mention that, independently from the Markov property, one can build representations of the ZF algebra  using the quantum group structure. To do so, one first observes that if $T(x)$ satisfies the RTT relation \eqref{rttg} and $A(x)$ obeys the ZF relation \eqref{eq:ZF}, then $T_{0q}(x)\,A_{0q'}(x)$ (where 0 denotes the auxiliary space and $q$ and $q'$ are two independent quantum spaces) also obeys the ZF relation:
\begin{equation}
 R_{12}(x_1/x_2) \,T_{1q}(x_1)\,A_{1q'}(x_1)\, T_{2q}(x_2)\,A_{2q'}(x_2)
 =T_{2q}(x_2)\,A_{2q'}(x_2)\,T_{1q}(x_1)\, A_{1q'}(x_1)\;.
\end{equation}
The construction is valid at the algebraic level.
  Thus, starting from a representation of the ZF algebra, one can build new ones using quantum group representations.  An example of such construction is the relation \eqref{eq:T-A}, where we have used the scalar representation \eqref{eq:marmr} for the ZF algebra and kept the quantum group generators unrepresented.

\subsection{Ghoshal-Zamolodchikov relations \label{sec:GZ}}

The Zamolodchikov relation \eqref{eq:ZF} allows one to deal with the bulk part of an integrable quantum field theory. 
To include boundaries which preserve integrability, an additional relation, called Ghoshal-Zamolodchikov (or GZ to shorten), is needed \cite{GZ}. It is given by
\begin{equation}\label{eq:GZ}
 \llangle \Omega|\Big(  K(x) A(1/x)-A(x) \Big)=0\quad,\qquad \Big(\bar K(x)A(1/x)-A(x) \Big) |\Omega \rrangle=0\;,
\end{equation}
where $|\Omega \rrangle$ is the vacuum state of the quantum field theory.

The consistency of these conditions is given by the reflection equation \eqref{eq:re} for the GZ relation on the left vector $ \llangle \Omega|$
and by the reflection equation \eqref{eq:re_bis} for the GZ relation on the right vector $|\Omega \rrangle$. 
There are for instance two ways of computing $\llangle \Omega| A_2(x_2)A_1(x_1)$
\begin{eqnarray}
 \llangle \Omega| A_2(x_2)A_1(x_1)&=&R_{12}(\frac{x_1}{x_2})\ \llangle \Omega| A_1(x_1)A_2(x_2)
= R_{12}(\frac{x_1}{x_2})K_1(x_1)\ \llangle \Omega| A_1(\frac1{x_1})A_2(x_2)\nonumber \\
 &=&R_{12}(\frac{x_1}{x_2})K_1(x_1)R_{21}(x_2x_1)\ \llangle \Omega| A_2(x_2)A_1(\frac1{x_1})\nonumber \\
 &=&R_{12}(\frac{x_1}{x_2})K_1(x_1)R_{21}(x_1x_2)K_2(x_2)\ \llangle \Omega| A_2(\frac1{x_2})A_1(\frac1{x_1}),
\end{eqnarray}
or
\begin{eqnarray}
 \llangle \Omega| A_2(x_2)A_1(x_1)&=&K_2(x_2)\ \llangle \Omega| A_2(\frac1{x_2})A_1(x_1)
= K_2(x_2)R_{12}(x_1x_2)\ \llangle \Omega| A_1(x_1)A_2(\frac1{x_2})\nonumber \\
 &=&K_2(x_2)R_{12}(x_1x_2)K_1(x_1)\ \llangle \Omega| A_1(\frac1{x_1})A_2(\frac1{x_2})\nonumber \\
 &=&K_2(x_2)R_{12}(x_1x_2)K_1(x_1)R_{21}(\frac{x_1}{x_2})\ \llangle \Omega| A_2(\frac1{x_2})A_1(\frac1{x_1}).
\end{eqnarray}
Therefore the reflection equation \eqref{eq:re} is a sufficient condition 
to ensure the consistency of the GZ relation for $K(x)$.
A similar result holds for $\bar K(x)$.

Another consistency relation is ensured by the unitarity property \eqref{eq:unitary_K}: if one applies the GZ relation twice
\begin{equation}
 \llangle \Omega|A(x)=K(x)\ \llangle \Omega|A(1/x)=K(x)K(1/x)\ \llangle \Omega|A(x).
\end{equation}

Due to relation \eqref{eq:T-A} linking $A(x)$ satisfying Zamolodchikov algebra and $T(x)$ satisfying the FRT relation, 
there exists a nice way to 
write the GZ relation \eqref{eq:GZ}
\begin{eqnarray}
&& v^t(1/x)\llangle \Omega|~ T(1/x)^t K^t(x) {T^t(x)}^{-1}  =v^t(x)~\llangle \Omega|\ ,\\
 &&  T(x)^{-1}{\bar K}(x)T(1/x) ~ |\Omega \rrangle v(1/x)=|\Omega \rrangle v(x)\;.
\end{eqnarray}

\subsection{Examples \label{sec:eqg}}

As explained previously, the RTT relation \eqref{rttg} allows one to provide in a compact form 
all the  defining relations of a quantum group. Therefore, for each explicit $R$-matrix, one gets a different quantum group.
In this section, we define the different quantum groups associated to the R-matrices introduced in section \ref{sec:exint}.
To be brief and clear, we will overlook some subtleties in the presentation, focusing on the  
 the main ideas.

\paragraph{Quantum affine algebra associated to ASEP.}

We study here the quantum group associated to the $R$-matrix \eqref{eq:RASEP} of the ASEP: it is called the quantum enveloping algebra 
of the affine Lie algebra $\widehat{gl_2}$ (or quantum affine algebra) and denoted $U_q(\widehat{gl_2})$ \cite{drinfeld,jimbo}.
The algebra $U_q(\widehat{gl_2})$ is defined by\footnote{In fact, to be precise, we will define only a subalgebra of $U_q(\widehat{gl_2})$: the upper Borel.
However, only this subalgebra is needed for the following.}
\begin{eqnarray}
  R_{12}(\frac{z}{w})\ {\cal L}_{1}(z)\ {\cal L}_{2}(w) 
  &=& {\cal L}_{2}(w)\ {\cal L}_{1}(z)\ R_{12}(\frac{z}{w}),
  \label{rtt-uq}
 \end{eqnarray}
where $R$ is given by \eqref{eq:RASEP}. The generators are gathered as follows
\begin{equation}
{\cal L}(z)=\left( \begin{array}{cc} {\cal A}(z)& {\cal B}(z) \\[1.ex] {\cal C}(z) & {\cal D}(z) \end{array} \right)
\end{equation}
with $\displaystyle {\cal A}(z)=\sum_{n=0}^{+\infty} z^{n}\,  {\cal A}^{(n)}$,...
We emphasize that the $R$-matrix \eqref{eq:RASEP} used here is not the one used usually to define $U_q(\widehat{gl_2})$. 
However, there exists an isomorphism between the algebra defined by \eqref{rtt-uq} and $U_q(\widehat{gl_2})$. 

As explained above, the generating series (${\cal A}(z),...,\cD(z)$) can be truncated \textit{i.e.} an infinite number of generators can be put to 0. 
The simplest evaluation map in the case considered here is given by
\begin{eqnarray}\label{eq:emuq}
 {\cal L}(z)=z\left( \begin{array}{cc} 1& {\cal B}^{(1)} \\[1.ex] 0 & {\cal D}^{(1)} \end{array} \right)+
 \left( \begin{array}{cc} {\cal A}^{(0)}& 0 \\[1.ex] {\cal C}^{(0)} & 1 \end{array} \right)\;.
\end{eqnarray}
The four generators present in the previous relation generate the finite quantum algebra $U_q(gl_2)$.

\paragraph{Yangian associated to SSEP.}

We study here the quantum group associated to the $R$-matrix \eqref{eq:RSSEP} of the SSEP: it is called the Yangian of ${gl_2}$ 
and denoted $Y(gl_2)$ (for a review see \cite{MNO,molev}). 
It is defined by the relations
\be\label{rtt-Y}
R_{12}(u_1-u_2)\,T_1(u_1)\,T_2(u_2) = T_2(u_2)\,T_1(u_1)\,R_{12}(u_1-u_2)
\ee
where R is given by \eqref{eq:RSSEP} and
\be
T(u)=\left( \begin{array}{cc} t_{11}(u)& t_{12}(u) \\ t_{21}(u) & t_{22}(u) \end{array} \right)\qquad\text{with}\qquad
t_{ij}(u)=\delta_{ij}+\sum_{n=1}^\infty t_{ij}^{(n)}\,u^{-n}\,.
\ee
For the Yangian, there exists also an evaluation map given by
\be\label{eq:rey}
T(u)=\left( \begin{array}{cc} 1& 0 \\ 0 & 1 \end{array} \right)
+\frac{1}{u}\left( \begin{array}{cc} t_{11}^{(1)}& t_{12}^{(1)} \\ t_{21}^{(1)} & t_{22}^{(1)} \end{array} \right)
\,.
\ee
The four generators present in \eqref{eq:rey} generate the Lie algebra $gl_2$.

\section{Quantum group and Matrix ansatz \label{sec:link}}

\subsection{Zamolodchikov algebra and matrix ansatz}

The main result of this section is to show that the algebra elements $E$, $D$, $\overline{E}$, $\overline{D}$ used in the matrix ansatz 
(section \ref{matrix_ansatz}) can be generated 
with a vector $A(x)$ depending on a spectral parameter $x$ and satisfying the Zamolodchikov algebra \eqref{eq:ZF}. 
The relations
between these objects are in fact very simple: 
\begin{equation}\label{eq:ide}
A(1) = \left(\begin{array}{c}E\\D
 \end{array}\right), \quad A'(1) =\rho \left(\begin{array}{c} \overline{E}\\ \overline{D}
 \end{array}\right)\;,
\end{equation} 
where $.'$ stands for the derivation with respect to the spectral parameter $x$ and $\rho$ appears in \eqref{eq:jump}.
Indeed, derivating the Zamolodchikov relation \eqref{eq:ZF} w.r.t. $x_1$ and putting $x_1=x_2=1$, we get
\begin{equation}
  R'(1)A_1(1)A_2(1)+R(1)A'_1(1)A_2(1)=A_2(1)A'_1(1)\;.
\end{equation}
Then, multiplying on the right by $P$ (the permutation operator) and using regularity \eqref{eq:regularity} and local jump operator \eqref{eq:jump} properties, 
the previous relation may be written as follows
\begin{equation}
 w\ A_1(1)A_2(1)=\frac{1}{\rho}\left( A_1(1)A'_2(1)-A'_1(1)A_2(1) \right)\;.
\end{equation}
Finally, using the identification \eqref{eq:ide}, we recover exactly relation \eqref{eq:tel} which is the central relation for the matrix ansatz.
Let us emphasize that the Zamolodchikov algebra may contain more relations than relation \eqref{eq:tel}. These supplementary relations can be crucial 
when one wants to compute explicitly the steady state using uniquely the algebra.\\

At this point, let us summarize what we achieved: starting from the RTT presentation of the quantum group (section \ref{sec:frt}), we use the 
Markovian property \eqref{eq:marmr} of the $R$-matrix to get a Zamolodchikov algebra (section \ref{sec:ZF}) and, derivating this relation, we get 
finally the cornerstone relation \eqref{eq:tel} for the matrix ansatz (section \ref{matrix_ansatz}). Therefore, we provide 
a precise link between the algebra necessary for the 
matrix ansatz and the quantum group underlying the model and used previously to prove its integrability (section \ref{sec:integrability}) 
and in the context of the algebraic Bethe Ansatz. 
Let us mention \cite{SW} where a first attempt to link Zamolodchikov algebra and matrix ansatz has been made.

\subsection{GZ relation and action on boundaries}

The other relations needed for the matrix ansatz concern the actions of the operators $E$, $D$, $\overline{E}$, $\overline{D}$ on $|V\rrangle$ and $\llangle W|$, as
given in \eqref{eq:telB}.
Again there exists a nice link between these relations  and the GZ relations introduced in
the context of integrable quantum field theory (section \ref{sec:GZ}).

Indeed, derivating relations \eqref{eq:GZ} w.r.t. $x$ and taking $x=1$, one gets
 \begin{equation}\label{eq:GZd}
 \llangle \Omega|\Big(  K'(1) A(1)-K(1)A'(1)-A'(1) \Big)=0\quad,\qquad \Big(\bar K'(1)A(1)-\bar K(1)A'(1)-A'(1) \Big) |\Omega \rrangle=0\;.
\end{equation}
Then, by identifying $\llangle \Omega|=\llangle W|$ and $|\Omega \rrangle=|V\rrangle$ and by using regularity \eqref{eq:regularity_K} 
and local jump operator \eqref{eq:jump_K} properties for the $K$-matrix, we obtain the following relations
\begin{equation}
 \llangle W|\left(BA(1)-\frac{1}{\rho}A'(1)\right)=0, \quad \left(\overline{B} A(1)+\frac{1}{\rho}A'(1)\right) |V\rrangle=0.
\end{equation}
With the identification \eqref{eq:ide}, we recover exactly relations \eqref{eq:telB}.
Therefore, the GZ relation implies the relations \eqref{eq:telB} and may contain more relations.\\

At this stage we can make the following observation: let us introduce $C(x):=\langle 1 | A(x)={\cal X}_1(x)+{\cal X}_2(x)$ which allows us to define
\begin{equation}
 Z_L(x_1,\dots,x_L):=\llangle W|C(x_1)\dots C(x_L) |V \rrangle\;.
\end{equation}
This function is a generalisation of the normalization factor $Z_L$ of the probability distribution since 
$Z_L(1,\dots,1)=\llangle W|C(1)^L |V\rrangle=\llangle W|(E+D)^L |V\rrangle=Z_L$.
Using the $R$-matrix Markovian property on the 
ZF relation \eqref{eq:ZF} one can show that $[C(x_1),C(x_2)]=0$. In the same way using the $K$-matrix Markovian property   on the GZ relations
\eqref{eq:GZ} we get $\llangle W | C(x)=\llangle W | C(1/x)$ and $C(x)|V \rrangle=C(1/x)|V \rrangle$. Remark that in particular $\llangle W| C'(1)=0$
and $C'(1) |V\rrangle =0$. These relations imply that 
$Z_L(x_1,\dots,x_L)$ is a symmetric polynomial in the variables $x_1,1/x_1,\dots,x_L,1/x_L$. We also have that
$\partial Z_L/\partial x_i (1,\dots,1)=0$.\\

The last step to obtain the matrix ansatz stationary state \eqref{eq:O}, in the context of Zamolodchikov relation 
and GZ condition, is to prove the existence of $|V\rrangle$ and $\llangle W|$ such that $\llangle W|V\rrangle\neq 0$. 
For such a purpose, it is enough to give an explicit representation for all the generators and for $|V\rrangle$ and $\llangle W|$.
This task must be done case by case, and we illustrate it in section \ref{sec:8v}.

Let us mention here that in the Razumov-Stroganov conjecture
context, the groundstate of the ($\Delta = −1/2$) loop model in presence of inhomogeneities has been obtained using similar technique
to those exposed in this section: the groundstate was identified with a representation of the Zamloldchikov relations 
(or Ghoshal-Zamloldchikov relations in the open case). See for instance \cite{DFZJ} and \cite{Cant}.

\subsection{Examples}

\paragraph{ASEP model.} From the evaluation map \eqref{eq:emuq} for the quantum affine algebra $U_q(\widehat{gl_2})$ and 
using relation \eqref{eq:T-A}, we get the 
Zamolodchikov generators 
\begin{equation}
 A(x)=f(x)\left(\begin{array}{c}
             ax^2+x(a{\cal A}^{(0)} + b {\cal B}^{(1)})\\
             x(a{\cal C}^{(0)}+b{\cal D}^{(1)})+b
            \end{array}
\right)\;.
\end{equation}
Through the identification \eqref{eq:ide}, and choosing $a=b=1$ and $f(x)=\frac{1}{x}$,  one gets
\begin{eqnarray}
 E=1+{\cal A}^{(0)} +  {\cal B}^{(1)}\quad&,&\quad\overline{E}=q-1\\
 D={\cal C}^{(0)}+{\cal D}^{(1)}+1\quad&,&\quad\overline{D}=1-q
\end{eqnarray}
We recover the values of the operators $\overline{E}$ and $\overline{D}$ given in section \ref{sec:MAex} and 
obtain a realisation of the generators $E$ and $D$ in terms of the $U_q(gl_2)$ generators.

\paragraph{SSEP model.} The relation \eqref{eq:T-A} applied in the Yangian case,  with the evaluation map \eqref{eq:rey},
leads to 
\begin{equation}
 A(x)=f(x)\left(\begin{array}{c}
            a+\frac{1}{x}(at_{11}^{(1)}+b t_{12}^{(1)}) \\ 
            b+\frac{1}{x}(at_{21}^{(1)} +b t_{22}^{(1)}) 
            \end{array}
\right)\;.
\end{equation}
Choosing $a=-1$, $b=1$ and $f(x)=x$ and using the identification \eqref{eq:ide}, one gets
\begin{eqnarray}
 E=-1-t_{11}^{(1)}+ t_{12}^{(1)}\quad&,&\quad\overline{E}=-1\\
 D=-t_{21}^{(1)} + t_{22}^{(1)}+1\quad&,&\quad\overline{D}=1
\end{eqnarray}
We recover again the values of the operators $\overline{E}$ and $\overline{D}$ given  in section \ref{sec:MAex}  
and obtain this time a realisation of the generators $E$ and $D$ in terms of the $gl_2$ ones. 
We recover also the realisation of the Zamolodchikov algebra found in \cite{SW}.

\section{Inhomogeneous transfer matrix\label{sec:ti}}

For the well-known ASEP, SSEP and TASEP models presented in this paper, the matrix ansatz allows one to obtain an
eigenvector of $M$. In the context of  integrable models, one usually computes 
 the transfer matrix eigenvectors.
In this section, we show that the Zamolodchikov algebra allows us to construct a transfer matrix eigenvector 
which is a generalization of the stationary state. This transfer matrix eigenvector depends on $L$ parameters $\theta_j$ (called inhomogeneities),
and reduces to the steady state when all inhomogeneities are set to 1.

Let us add that to find an eigenvector for a transfer matrix with non-diagonal boundary is a difficult task 
(see section \eqref{sec:ba} for a more detailed discussion)
and is the first step to use the usual Bethe ansatz approaches.

\subsection{Eigenvector of the transfer matrix}

A generalization of the transfer matrix introduced in section \ref{transfer_matrix} can be obtained by adding inhomogeneity parameters
$\theta_1,\dots,\theta_L$ in the following way
\begin{equation}\label{eq:tri}
 t(x)=tr_0\big(\ \widetilde K_0(x)\ R_{0L}(\frac{x}{\theta_L})\dots R_{01}(\frac{x}{\theta_1})\ K_0(x)\ R_{10}(x\theta_1)\dots R_{L0}(x\theta_L)  \ \big)\;.
\end{equation}
This inhomogeneous transfer matrix \eqref{eq:tri} still commutes for different spectral parameters.
We show below that the following vector
 \begin{equation}\label{eq:Sin}
 |{\cal S}(\theta_1,\dots,\theta_L)\rangle=\llangle W| A_1(\theta_1)\dots A_L(\theta_L) |V\rrangle
\end{equation}
is an eigenvector of this inhomogeneous transfer matrix
when $x=\theta_i$ or $x=1/\theta_i$ with the eigenvalue equal to 1.
As already mentioned, the  vector \eqref{eq:Sin} is a generalisation of the stationary state \eqref{eq:O} given by the matrix ansatz 
\begin{equation}
 \steady=\frac{1}{Z_L}\ |{\cal S}(1,\dots,1)\rangle\;.
\end{equation}

One can get a nice expression for $t(x)$ by evaluating it at the particular points $x=\theta_1,\dots,\theta_L$ and
$x=1/\theta_1,\dots,1/\theta_L$. For instance one has
\begin{eqnarray}\label{eq:ttheta_i}
 t(\theta_i) &=& tr_0\Big(\widetilde K_0(\theta_i)\ R_{0L}(\frac{\theta_i}{\theta_L})\dots P_{0i}\dots R_{01}(\frac{\theta_i}{\theta_1})\ K_0(\theta_i)\ R_{10}(\theta_i\theta_1)\dots R_{L0}(\theta_i\theta_L) \Big) \\
 &=& R_{i,i-1}(\frac{\theta_i}{\theta_{i-1}}) \dots R_{i1}(\frac{\theta_i}{\theta_1})\ K_i(\theta_i)\  R_{1i}(\theta_i\theta_1) \dots R_{i-1,i}(\theta_i\theta_{i-1})\ \times \\
 & & tr_0\Big( \widetilde K_0(\theta_i)\ R_{0L}(\frac{\theta_i}{\theta_L})\dots  R_{0,i+1}(\frac{\theta_i}{\theta_{i+1}})\ R_{0i}(\theta_i^2)\ R_{i+1,i}(\theta_i\theta_{i+1})\dots R_{Li}(\theta_i\theta_L)\ P_{0i} \Big)\nonumber \\
 &=& R_{i,i-1}(\frac{\theta_i}{\theta_{i-1}}) \dots R_{i1}(\frac{\theta_i}{\theta_1})\ K_i(\theta_i)\  R_{1i}(\theta_i\theta_1) \dots R_{i-1,i}(\theta_i\theta_{i-1})\ \times \\
 & & R_{i+1,i}(\theta_i\theta_{i+1}) \dots R_{L,i}(\theta_i\theta_{L}) \  
 tr_0\Big( \widetilde K_0(\theta_i)\  R_{0i}(\theta_i^2)\ P_{0i} \Big) \ 
 R_{i,L}(\frac{\theta_i}{\theta_L}) \dots R_{i,i+1}(\frac{\theta_i}{\theta_{i+1}})\nonumber \\
 &=& R_{i,i-1}(\frac{\theta_i}{\theta_{i-1}}) \dots R_{i1}(\frac{\theta_i}{\theta_1})\ K_i(\theta_i)\  R_{1i}(\theta_i\theta_1) \dots R_{i-1,i}(\theta_i\theta_{i-1})\ \times \nonumber \\
 & & R_{i+1,i}(\theta_i\theta_{i+1}) \dots R_{L,i}(\theta_i\theta_{L}) \  {\bar K}(\frac1{\theta_i})  
  \ R_{i,L}(\frac{\theta_i}{\theta_L}) \dots R_{i,i+1}(\frac{\theta_i}{\theta_{i+1}}).\label{eq:ti}
\end{eqnarray}
The first equality uses the regularity property \eqref{eq:regularity}.
The second equality is obtained by pushing $P_{0i}$ to the right (which permutes $0$ and $i$ in all the indices in the R-matrices and $K$-matrix
$P_{0i}$ went trough). Then one remarks that the R-matrices and $K$-matrix whose indices are between $1$ and $i$ can be moved out of the trace to the left
because they commute with the operators with indices between $i+1$ and $L$ or $0$. One then get the third equality using $L-i$ times the Yang-Baxter
equation
\begin{equation}
 R_{0j}(\theta_i/\theta_j)\ R_{0i}(\theta_i^2)\ R_{ji}(\theta_i\theta_j)=R_{ji}(\theta_i\theta_j) \ R_{0i}(\theta_i^2) \ R_{0j}(\theta_i/\theta_j),
\end{equation}
with $j=i+1, \dots, L$. The last relation is obtained using definition \eqref{eq:Kbar} of $\bar K$.

Using expression \eqref{eq:ti} of $t(\theta_i)$, one shows that
\begin{equation} \label{eq:vpi}
  t(\theta_i)\ |{\cal S}(\theta_1,\dots,\theta_L)\rangle=|{\cal S}(\theta_1,\dots,\theta_L)\rangle\;.
\end{equation}
To prove it, one picks up the generator $A_i(\theta_i)$ and moves it completely to the right using the Zamolodchikov relation \eqref{eq:ZF}. Then one can reflect $A_i(\theta_i)$ against 
the left vector $|V\rrangle$ using equation \eqref{eq:GZ}. After this reflection, one can push $A_i(1/\theta_i)$ completely to the left using 
$L-1$ times equation \eqref{eq:ZF} and reflect it against $\llangle W|$ with \eqref{eq:GZ}. Finally, one can complete the cycle by moving back $A_i(\theta_i)$ to its initial position. 
One can show using the same arguments that 
\begin{equation} \label{eq:vpibis}
  t(1/\theta_i)\ |{\cal S}(\theta_1,\dots,\theta_L)\rangle=|{\cal S}(\theta_1,\dots,\theta_L)\rangle\;.
\end{equation}
Similar results have been obtained in \cite{SW} but for a different transfer matrix.

Let us emphasize that relations \eqref{eq:vpi} and \eqref{eq:vpibis} have been obtained without assuming a specific form for the $R$-matrix.
Thus, they are valid for any model, provided the $R$-matrix and the $K$-matrices obey the relations depicted in sections \ref{sec:Rmat}, \ref{sec:ZF}  and \ref{sec:GZ}. 
Moreover, one can always normalize the $R$-matrix  in such a way that the corresponding transfer matrix becomes a polynomial in the spectral parameter $x$. Then, knowing relations \eqref{eq:vpi} and \eqref{eq:vpibis}, plus some crossing symmetry for $t(x)$ and its $x\to\infty$ behavior, is enough to deduce (through interpolation) that the vector \eqref{eq:Sin} is an eigenvector of $t(x)$. However, to do this final step, one needs to know the precise form of $R(x)$. 
 Thus it can be done case by case only: below, we illustrate this final step for the case of ASEP and SSEP models.

\subsection{Examples}
We  show that, for the SSEP and ASEP models, one gets  
\bea \label{eq:vpg}
 t(x)\ |{\cal S}(\theta_1,\dots,\theta_L)\rangle &=&
 \lambda(x;\theta_1,\dots,\theta_L)\  |{\cal S}(\theta_1,\dots,\theta_L)\rangle
\eea 
where the eigenvalue $\lambda(x;\theta_1,\dots,\theta_L)$ is given by
\bea\label{eq:LS}
 \lambda(x;\theta_1,\dots,\theta_L) &=& 1+\frac{(x+1)(\delta+\beta)-1}{x(\delta+\beta)+1}\frac{(x+1)(\alpha+\gamma)-1}{x(\alpha+\gamma)+1}
 \frac{x}{x+1}\prod_{i=1}^{L}\frac{x^2-\theta_i^2}{(x+1)^2-\theta_i^2},\qquad
\eea
for the SSEP model, while for the ASEP model, one has
\bea \label{eq:LA}
\lambda(x;\theta_1,\dots,\theta_L)&=& 1+\frac{(qx-1)(\delta+qx\beta)+qx(1-q)}{(x-1)(x\delta+\beta)+x(q-1)}\ \frac{(qx-1)(qx\alpha+\gamma)+qx(1-q)}{(x-1)(\alpha+x\gamma)+x(q-1)}
\nonumber \\ 
 & &\quad \times q^{L-1}\frac{x^2-1}{q^2x^2-1}\ \prod_{i=1}^{L}\frac{(x-\theta_i)(x\theta_i-1)}{(qx-\theta_i)(qx\theta_i-1)}.
\eea
The proof is given below, first for the SSEP model, and then for the ASEP one.

\paragraph{SSEP model.} 
To prove relation \eqref{eq:vpg}, we use firstly the following properties of the $R$ and $K$ matrices
\begin{itemize}
 \item PT-symmetry: $R_{12}^{t_1t_2}(x)=R_{21}(x)=R_{12}(x)$,
 \item Crossing symmetry: 
 \bea 
 R_{12}(x)&=& x/(x+1)V_1R_{12}^{t_2}(-x-1)V_1^{-1}\quad \mbox{where} \quad V=\left( \begin{array}{cc}
            0&1\\
            -1&0
           \end{array}
 \right),
\\
   \widetilde K(x)&=&-\frac{2x+1}{2(x+1)}\frac{(x+1)(\delta+\beta)-1}{x(\delta+\beta)+1}V \left. K^t(-x-1) \right\vert_{\alpha\rightarrow\delta,\gamma\rightarrow\beta} V^{-1},
 \eea
 \item Duality relation: $\left. K_1(x)\right\vert_{\alpha\rightarrow\delta,\gamma\rightarrow\beta}=tr_0(\widetilde K_0(x)R_{01}(2x)P_{01})$,
\end{itemize}
to show the relation (following the method developed in \cite{TQ1,TQ2,TQ3})
\begin{eqnarray} \label{eq:sym_t}
 t(x)&=&
 \Big(\lambda(x;\theta_1,\dots,\theta_L)-1\Big)\; t(-x-1),
\end{eqnarray}
where $\lambda(x;\theta_1,\dots,\theta_L)$ is given in \eqref{eq:LS}.
Secondly, let us define 
$\widetilde t(x)=(x(\delta+\beta)+1)(x(\alpha+\gamma)+1)(x+1)\prod_{i=1}^{L}((x+1)^2-\theta_i^2)\,t(x)$. 
It is easy to check that the entries of $\widetilde t$ are polynomials in $x$ of degree less than $2L+3$. 
Then, using \eqref{eq:vpi}, \eqref{eq:vpibis} and \eqref{eq:sym_t} we see
that $|{\cal S}(\theta_1,\dots,\theta_L)\rangle$ is an eigenvector of $\widetilde t(x)$ for $4L$ different values $x=\theta_i,-\theta_i,\theta_i-1,-\theta_i-1$.
Hence we get \eqref{eq:LS} through interpolation in $x$.

\paragraph{ASEP model.}
Similarly to the SSEP case, to prove relation \eqref{eq:LA}, one needs the following symmetry properties of the $R$
and $K$ matrices:
\begin{itemize}
 \item T-symmetry: $R_{12}^{t_1t_2}(x)=M_1R_{21}(x)M_2^{-1}$,
 \item P-symmetry: $R_{21}(x)=V_1V_2R_{12}(x)V_1^{-1}V_2^{-1}=W_1W_2R_{12}W_1^{-1}W_2^{-1}$,
 \item $Z_2$-symmetry: $R_{12}(x)=M_1M_2R_{12}(x)M_1^{-1}M_2^{-1}$,
 \item Crossing symmetry: 
 \bea
 R_{12}(x)&=& (x-1)/(qx-1)M_1V_1R_{12}^{t_2}(1/qx)V_1^{-1},
 \\
V^{t} \widetilde K^{t}\left(\frac{1}{qx}\right)V^{-1}&=&\frac{qx^2-1}{x^2-1}\frac{(x-1)(x\delta+\beta)+x(q-1)}{(qx-1)(qx\beta+\delta)+qx(1-q)} MW
  \left.K(x)\right\vert_{\atopn{\alpha\rightarrow\beta}{\gamma\rightarrow\delta}} W^{-1},\qquad
\eea
 \item Duality relation: $ W_1\left. K_1(x)\right\vert_{\atopn{\alpha\rightarrow\beta}{\gamma\rightarrow\delta}}W_1^{-1}=tr_0\big(\widetilde K_0(x)R_{01}(2x)P_{01}\big)$,
\end{itemize}
where
\be
 V=\left( \begin{array}{cc}  0&-1\\  q&0 \end{array} \right)\,,\
 W=\left( \begin{array}{cc} 0&1\\ 1&0 \end{array} \right) 
 \mbox{ and }
  M=\left( \begin{array}{cc} q&0\\  0&1  \end{array} \right)\;.
\ee

Using these symmetry properties, we can show the following relation
\begin{eqnarray}\label{eq:cra}
 t(x) &=& \Big(\lambda(x;\theta_1,\dots,\theta_L)-1\Big)\; t(1/qx),
\end{eqnarray}
where $\lambda(x;\theta_1,\dots,\theta_L)$ is given in \eqref{eq:LA}.
As we did previously for the SSEP model, relation \eqref{eq:cra} allows us to prove that $|{\cal S}(\theta_1,\dots,\theta_L)\rangle$
is eigenvector for the inhomogeneous ASEP transfer matrix with the eigenvalue \eqref{eq:LA}.

The eigenvalue of the homogeneous transfer matrix is easily obtained by taking the limit $\theta_i=0$ in \eqref{eq:LS} for the SSEP model
and $\theta_i=1$ in \eqref{eq:LA} for the ASEP model.

\paragraph{Discrete time Markov process}
For the SSEP and ASEP models, one can show in the same way that the row vector $\langle 1|$ with all entries equal to $1$ is a left eigenvector of $t(x)$
associated to the eigenvalue $\lambda(x;\theta_1,\dots,\theta_L)$. Hence the matrix $t(x)$ should describe, up to a normalization, a discrete time
Markov process. It would be very interesting to understand precisely what this process is. It has been done for the transfer matrix of the periodic  TASEP model in \cite{tMarkov}.

\subsection{Comparison with the Bethe ansatz approaches \label{sec:ba}}

The computation of the eigenvectors and eigenvalues of the inhomogeneous transfer matrix \eqref{eq:tri}
for non-diagonal $K$-matrices has attracted a lot of attentions since a direct use of the usual methods based on the Bethe ansatz fails.
We summarize here the results obtained previously and compare them with the results obtained by matrix ansatz.

\paragraph{SSEP model.}

For the SSEP (or XXX) model, one can conjugate the transfer matrix to simplify the $K$-matrices:
\begin{eqnarray}
 t^s(x)&=&\Gamma_1^{-1}\dots \Gamma_L^{-1} t(x) \Gamma_1\dots \Gamma_L\\
 &=& tr_0\Big( \widetilde D_0(x)\ R_{0L}(x/\theta_L)\dots R_{01}(x/\theta_1)\ D_0(x)\ R_{10}(x\theta_1)\dots R_{L0}(x\theta_L) \Big)\;,\label{eq:eq1}\\
\mbox{where}&& \widetilde D(x)=\Gamma^{-1} \widetilde  K(x) \Gamma \mb{and} D(x)= \Gamma^{-1} K(x)\Gamma. 
\end{eqnarray}
To get \eqref{eq:eq1}, we have used the invariance property of the $R$-matrix 
$\Gamma_1\Gamma_2R_{12}(x)=R_{12}(x)\Gamma_1\Gamma_2$, valid for any $\Gamma$ and in the XXX case, and the cyclicity of the trace (see \cite{AACDFR2,GM} for a detailed proof).
For the Markovian boundaries \eqref{eq:Kds} and \eqref{eq:Kts}, we get for 
$\Gamma=\left( \begin {array}{cc} 
-1&\beta\\ 
1&\delta
\end {array} \right)$
\bea
 \widetilde D(x) &=& \frac{2x+1}{2(x+1)(x(\delta+\beta)+1)}\left( \begin {array}{cc} 
-(x+1)(\beta+\delta)+1&0\\ 
0&(x+1)(\beta+\delta)+1
\end {array} \right)\,,
\\[1.2ex]
D(x) &=& \left( \begin {array}{cc} 
-\frac{x(\alpha+\gamma)-1}{x(\alpha+\gamma)+1}&\frac{2x(\alpha\beta-\delta\gamma)}{x(\alpha+\gamma)+1}\\ 
0&1
\end {array} \right).
\eea
Remark that we did not impose any constraint on the boundary parameters (but the Markovian property).

When $\alpha\beta-\delta\gamma=0$, both boundaries can be diagonalized simultaneously and we recover the case 
already solved by
the algebraic Bethe ansatz  in \cite{sklyanin}. Remark that this last relation just corresponds
to the case where the matrix ansatz does not work (see e.g. \cite{ER}).

The case when one boundary is diagonal and the other one is triangular has been also solved by the algebraic Bethe ansatz in \cite{MMR}
(or it can be seen as a limit of the case with both boundaries triangular treated in \cite{BCR}). 
This method provides the eigenvalues and the eigenvectors of $t^s(x)$. The eigenvalues are not modified in comparison to the diagonal case.
Although they are well-known, we give them in the SSEP notations (with the normalisations of the $R$ and $K$
matrices used in the present paper) 
to easily compare them with the matrix ansatz 
\begin{eqnarray}
 \lambda^s(x)&=&\frac{1-(\beta+\delta)x}{1+(\beta+\delta)x}\ \frac{1-(\alpha+\gamma)x}{1+(\alpha+\gamma)x}\prod_{k=1}^M f(x,x_k)\\
 &&+\frac{x}{x+1}\frac{1+(\beta+\delta)(x+1)}{1+(\beta+\delta)x}\ \frac{1+(\alpha+\gamma)(x+1)}{1+(\alpha+\gamma)x}\prod_{j=1}^L\frac{x^2-\theta_j^2}{(x+1)^2-\theta_j^2}\ \prod_{k=1}^M h(x,x_k)\,, \nonumber
\end{eqnarray}
where $f(x,y)=\frac{(x-y-1)(x+y)}{(x+y+1)(x-y)}$ and $h(x,y)=\frac{(x-y+1)(x+y+2)}{(x-y)(x+y+1)}$.
As usual, one has introduced the integer $M$ chosen between $0$ and $L$ which corresponds to the maximum number of spin down in the eigenvectors. 
The parameters $x_1,x_2,\dots,x_M$ are called Bethe roots and must satisfied the Bethe equations
\begin{eqnarray}
 \frac{1-(\beta+\delta)x_k}{1+(\beta+\delta)(x_k+1)}\ \frac{1-(\alpha+\gamma)x_k}{1+(\alpha+\gamma)(x_k+1)}\prod_{j=1}^L\frac{(x_k+1)^2-\theta_j^2}{x_k^2-\theta_j^2}
 =\prod_{j\neq k} \frac{h(x_k,x_j)}{f(x_k,x_j)}\,.
\end{eqnarray}
Let us remark that $\lambda^s(x)$ satisfies the symmetry relation 
\begin{eqnarray} 
 \lambda^s(x)+\lambda^s(-x-1)&=&
 \lambda^s(x)\; \lambda^s(-x-1),
\end{eqnarray}
that can be deduced from \eqref{eq:sym_t}.

The usual claim in the Bethe ansatz approach is that one gets all the eigenvalues of the transfer matrix $t^s(x)$ and then of $t(x)$ 
(since they share the same spectrum). Therefore, the eigenvalue \eqref{eq:LS} obtained by the matrix ansatz 
must be a particular solution of the Bethe ansatz approach. 
By a direct computation, we can show that
\be\label{eq:scal}
 \langle - | t^s(x) |-\rangle = \lambda(x;\theta_1,\dots,\theta_L) 
 \mb{where} |-\rangle=\left(\begin{array}{c}0\\1\end{array}\right)^{\otimes L}
\ee
and $\lambda(x;\theta_1,\dots,\theta_L)$ is the eigenvalue \eqref{eq:LS}.
Let us remark that $|-\rangle $ is not an eigenvector of $t^s(x)$ because $D(x)$ is not diagonal, but relation \eqref{eq:scal}
 suggests that to find \eqref{eq:LS}, we must choose $M=L$. We confirmed it by studying the cases $L=M=1$ analytically and 
$L=M=2$ numerically for particular choices of parameters.
Therefore, it seems that the matrix ansatz provides the more complicated eigenvector (\textit{i.e.} with the maximum 
number of Bethe roots) in the 
context of the algebraic Bethe ansatz. It is an interesting open problem to prove this statement and to compare also the eigenvectors.

\paragraph{ASEP model.} Let us first remind what is known for the XXZ model, 
where the situation is more complicated than for the XXX one. Indeed, in this case, the $R$-matrix is not  invariant anymore, so that we cannot transform the transfer matrix as in the XXX case. However, a more complicated transformation
can be performed on the transfer matrix. Then, the algebraic Bethe ansatz can be used if there is a constraint between the parameters of both boundaries \cite{CYSW03}. 
For the generic boundaries \eqref{soluBg}, this constraint reads
 (see also \cite{DE}\footnote{Only the two first factors appears in this paper 
since they considered only positive parameters and therefore the last two are irrelevant.} and \cite{CRS2})
\begin{equation}\label{eq:cons}
(\alpha\beta e^s-\gamma\delta q^{L-1-M})(e^s-q^{L-1-M})(\beta e^s+\delta q^{L-1-M})(\alpha e^s+\gamma q^{L-1-M})=0 
\end{equation}
where $M=0,1,\dots,L-1$ is the number of Bethe roots. 
These constraints appeared already in \cite{Nepo,murN} where the eigenvalues and the Bethe equations are computed by functional methods.
Let us also mention that functional methods have been also developed when $q$ is a root of unity \cite{Nepoqr,murNS}.
The coordinate Bethe ansatz has been also generalized in \cite{DS,CRS2} to get the eigenvectors and eigenvalues and give a nice interpretation of $M$.
Indeed, for a given $M$ and if the constraint \eqref{eq:cons} is satisfied, a basis
is constructed in \cite{DS,CRS2} where the Markov matrix can be written as the following tridiagonal bloc
$K$-matrix 
\begin{equation}\label{eq:madi}
 W=\left(\begin{array}{c c c c c c c c}
       n=0 & * &  & \\
       *  & n=1 & * & & \\
       & * & n=2 & * &  &  \\
       & & &\ddots\\
       &&&*&n=M & * & \\
       &&&&0&n=M+1 & * & \\
       &&&&&&\ddots\\
       &&&&&&*&n=L
       \end{array}
\right)\;.
\end{equation}
In the matrix \eqref{eq:madi}, the value of $n$ is the number of excitations in the corresponding square sub-matrices of dimension 
$\left(\begin{array}{c} L\\n \end{array}\right) \times \left(\begin{array}{c} L\\n \end{array}\right)$, $*$ stands for non-vanishing submatrices 
and all the other non-displayed submatrices  are 0. Because of the vanishing of the submatrix on the left of the block ``$n=M+1$'', the action
of $W$ on states with $n\leq M$ provides again states with $n\leq M$. Therefore, $M$ is interpreted as the maximum number of excitations for the
eigenvectors obtained in \cite{CYSW03,CRS2}.
With this interpretation, we conclude immediately that some eigenvectors are missing.
Numerical evidences of this fact have been given in \cite{NR}. 
Therefore, to complete the spectrum, a second set of Bethe equations has been found \cite{NR,NYZ}.
The left eigenvectors corresponding to these eigenvalues are easily computed for example by the generalized coordinate Bethe ansatz introduced in \cite{DS,CRS2}
using the canonical dual of the bases where $W$ takes the form \eqref{eq:madi}. This time the states contain at least $M+1$ excitations (it easy to check 
that this sector with at least $M+1$ excitations is stable by the action of $W$ on left vector) and we use the symmetry hole-particle.
The corresponding right eigenvectors are given by a generalisation of the matrix ansatz using the idea of the coordinate Bethe ansatz \cite{CRS1}.
For this case, the algebraic Bethe ansatz has been studied in \cite{YZ}: a comparison with the matrix ansatz would be interesting.\\

Let us come back to the ASEP case. In this case, the integrable boundaries must be markovian (\textit{i.e.} $e^s=1$) and the 
constraint \eqref{eq:cons} is satisfied for $M=L-1$. Therefore, the Bethe approaches provide all the eigenvalues and eigenvectors except one, the steady state
given by the matrix ansatz \cite{DEHP,sandow}. In \cite{DE}, the Bethe equations 
were used to compute the spectral gap and determine how states 
approach the steady state. 
 
As explained in section \ref{sec:eax}, it would be also interesting to study the more general integrable boundary with $e^s\neq 1$ 
to get the current fluctuations.
One can use the above results when the constraint \eqref{eq:cons} is satisfied.
For the generic case, progress has been made recently. The matrix ansatz has been developed in \cite{GLMV,laza}. The algebraic Bethe 
ansatz in the case of the SSEP has been generalized in \cite{BC13} based on the eigenvalues and Bethe equations found in \cite{CYSW,nepoXXX}. 
A lot of other methods have been also introduced: the Onsager's approach \cite{BK}, 
the functional method \cite{G},
the separation of variables method for SSEP model \cite{FSW} and for ASEP model \cite{N,FKN,KMN}, the TQ relation \cite{LP} and the 
inhomogeneous TQ relation \cite{CYSW13,CYSW14}.

\section{Example: a reaction-diffusion model \label{sec:8v}}

The aim of this section is to illustrate the use of  relations \eqref{eq:ZF} and \eqref{eq:GZ} to find the stationary state, on a model for which one needs an algebra
 more involved  than in the previous cases. As we will see, this algebra will be generated by 
three elements (instead of two in the previous cases) and the entries of $A'(1)$ will not be central in the algebra anymore.

\subsection{The model \label{sec:model8}}

\begin{figure}[htb]
\begin{center}
 \begin{tikzpicture}[scale=0.7]
\draw (-2,0) -- (12,0) ;
\foreach \i in {-2,-1,...,12}
{\draw (\i,0) -- (\i,0.4) ;}
\draw[->,thick] (-2.4,0.9) arc (180:0:0.4) ; \node at (-2.,1.8) [] {$\alpha$};
\draw[->,thick] (-1.6,-0.1) arc (0:-180:0.4) ; \node at (-2.,-0.8) [] {$\gamma$};
\draw  (1.5,0.5) circle (0.3) [fill,circle] {};
\draw  (4.5,0.5) circle (0.3) [fill,circle] {};
\draw  (5.5,0.5) circle (0.3) [fill,circle] {};
\draw  (8.5,3.1) circle (0.3) [fill,circle] {};
\draw  (9.5,3.1) circle (0.3) [fill,circle] {};
\draw[->,thick] (1.4,1) arc (0:180:0.4); \node at (1.,1.8) [] {$\kappa^2$};
\draw[->,thick] (1.6,1) arc (180:0:0.4); \node at (2.,1.8) [] {$\kappa^2$};
\node at (5,1.1) [rotate=-90] {$\Big{\{}$};
\draw[->,thick] (5,1.3) -- (5,2.8); \node at (5.3,2.2) [] {$1$};
\node at (9,2.5) [rotate=90] {$\Big{\{}$};
\draw[->,thick] (9,2.3) -- (9,0.8); \node at (9.3,2) [] {$1$};
\draw[->,thick] (11.6,1) arc (180:0:0.4) ; \node at (12.,1.8) [] {$\beta$};
\draw[->,thick] (12.4,-0.1) arc (0:-180:0.4) ; \node at (12.,-0.8) [] {$\delta$};
 \end{tikzpicture}
 \end{center}
 \caption{Reaction diffusion model.}
 \label{fig:rdm}
\end{figure}
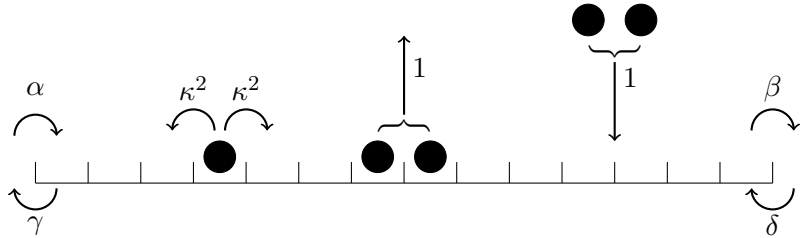

The stochastic process we want to study is a reaction-diffusion model. The particles move to the left and to the right with rate $\kappa^2$ 
and with a simple exclusion principle. They are injected and extracted on the boundaries with the same rates than the ASEP model. 
Moreover, two particles can be annihilated (``evaporate from the lattice'') when they are neighbours or created (``condensate'' on the lattice) 
on two adjacent empty sites with rate 1 in both cases. The dynamics is summarized in  figure \ref{fig:rdm}.

For this process, the local jump operators are given explicitly by 
\begin{equation} \label{eq:localw8vertex}
B =\left( \begin {array}{cc} 
-\alpha&\gamma\\ 
\alpha&-\gamma
\end {array} \right)\quad ,\qquad  w=\left( \begin {array}{cccc} 
-1&0&0&1\\ 
0&-\kappa^2&\kappa^2&0\\
0&\kappa^2&-\kappa^2&0\\
1&0&0&-1
\end {array} \right)
\quad\text{,}\qquad
\overline{B} =\left( \begin {array}{cc} 
-\delta&\beta\\ 
\delta&-\beta
\end {array} \right)\;.
\end{equation}
The use of relations \eqref{eq:tele} and \eqref{eq:telB}  to compute the stationary state shows that $\overline{E}$ and $\overline{D}$ 
cannot be scalar. Indeed, if $\overline{E}$ and $\overline{D}$ were scalars, the words of two letters would vanish (for instance).
Therefore, we propose in the following a more complicated algebra deduced from the Zamolodchikov algebra that allows us to compute any word.

\subsection{The $R$ and $K$ matrices}

The first step is again to compute the $R$-matrix and the $K$-matrix associated to the considered model.
 The $R$-matrix associated to the reaction-diffusion model is given by
\begin{equation}\label{eq:R8}
R(x)=\left( \begin {array}{cccc} 
\displaystyle{\frac{\kappa(x+1)}{\kappa(x+1)+x-1}}&0&0&\displaystyle{\frac{x-1}{\kappa(x+1)+x-1}}\\ 
0&\displaystyle{\frac{\kappa(x-1)}{\kappa(x-1)+x+1}}&\displaystyle{\frac{x+1}{\kappa(x-1)+x+1}}&0\\
0&\displaystyle{\frac{x+1}{\kappa(x-1)+x+1}}&\displaystyle{\frac{\kappa(x-1)}{\kappa(x-1)+x+1}}&0\\
\displaystyle{\frac{x-1}{\kappa(x+1)+x-1}}&0&0&\displaystyle{\frac{\kappa(x+1)}{\kappa(x+1)+x-1}}
\end {array} \right) .
\end{equation}
It satisfies the Yang-Baxter equation, the unitarity, regularity, local jump  (with $\rho=1/(2\kappa)$ and $w$ given in \eqref{eq:localw8vertex}) 
and crossing unitarity properties with 
$$
U=1\,, \quad Q=(\kappa+1)^2/(\kappa-1)^2 \mb{and}
 \lambda(x)=\frac{(x^2-1)(x(\kappa+1)^2+(\kappa-1)^2)(x(\kappa+1)^2-(\kappa-1)^2)}{(x(\kappa+1)+\kappa-1)^2(x(\kappa+1)-(\kappa-1))^2}\,.
$$

We introduce a solution of the reflection equation \eqref{eq:re} associated to this $R$-matrix:
\begin{equation}
K(x)=\left( \begin {array}{cc} 
\displaystyle{\frac{(x^2+1)((x^2-1)(\gamma-\alpha)+4x\kappa)}{2x((x^2-1)(\alpha+\gamma)+2\kappa(x^2+1))}}&\displaystyle{\frac{(x^2-1)((x^2+1)(\gamma-\alpha)+2x(\alpha+\gamma))}{2x((x^2-1)(\alpha+\gamma)+2\kappa(x^2+1))}}\\ 
\displaystyle{-\frac{(x^2-1)((x^2+1)(\gamma-\alpha)-2x(\alpha+\gamma)}{2x((x^2-1)(\alpha+\gamma)+2\kappa(x^2+1))}}&\displaystyle{-\frac{(x^2+1)((x^2-1)(\gamma-\alpha)-4x\kappa)}{2x((x^2-1)(\alpha+\gamma)+2\kappa(x^2+1))}}
\end {array} \right),
\end{equation}
and a solution of the reflection equation \eqref{eq:re_bis}
\begin{equation}
\bar K(x)= \left( \begin {array}{cc} 
\displaystyle{\frac{(x^2+1)((x^2-1)(\delta-\beta)+4x\kappa)}{2x(-(x^2-1)(\delta+\beta)+2\kappa(x^2+1))}}&\displaystyle{\frac{(x^2-1)((x^2+1)(\delta-\beta)-2x(\delta+\beta))}{2x(-(x^2-1)(\delta+\beta)+2\kappa(x^2+1))}}\\ 
\displaystyle{-\frac{(x^2-1)((x^2+1)(\delta-\beta)+2x(\delta+\beta)}{2x(-(x^2-1)(\delta+\beta)+2\kappa(x^2+1))}}&\displaystyle{-\frac{(x^2+1)((x^2-1)(\delta-\beta)-4x\kappa)}{2x(-(x^2-1)(\delta+\beta)+2\kappa(x^2+1))}}
\end {array} \right).
\end{equation}
These matrices satisfy the unitarity and regularity relations. As explained previously, 
they encode the boundaries \eqref{eq:localw8vertex} through their derivatives (boundary jump operator property)
\begin{equation}
 \kappa\ K'(1)=B \mb{and} -\kappa\ \bar K'(1)=\overline{B}\;.
\end{equation}
Using these matrices and the transfer matrix \eqref{eq:tb}, we prove that this model is integrable.

\subsection{Construction of the algebra}

We take the following expansion for the vector $A(x)$ satisfying the Zamolodchikov relation \eqref{eq:ZF}:
\begin{equation}
 A(x)= \left( \begin{array}{c}
   \displaystyle{G_1x+G_2+\frac{G_3}{x}}\\[2ex]
   \displaystyle{-G_1x+G_2-\frac{G_3}{x}}            
              \end{array} \right).
\end{equation}
We get the exchange relations between $G_1,G_2$ and $G_3$ using the relation \eqref{eq:ZF}. 
These exchange relations read 
\begin{equation} \label{eq:com8vertex}
 \left\{ \begin{aligned}
          & \phi\ G_1G_2=G_2G_1 \\
          & G_1G_3=G_3G_1 \\
          & \phi\ G_2G_3=G_3G_2
         \end{aligned}
 \right.\qquad\text{with}\qquad \phi=\frac{\kappa-1}{\kappa+1}\,.
\end{equation}
Using the GZ relation \eqref{eq:GZ}, we get the relations on the boundaries
\begin{equation} \label{eq:bords8vertex}
 \left\{ \begin{aligned}
          & \llangle W |\big( G_1-c\;G_2-a\;G_3 \big)=0\,, \\
          &\big( G_3-b\;G_1-d\;G_2 \big) | V \rrangle =0
         \end{aligned}
 \right.
 \mb{with}
 \left\{ \begin{aligned}
& a=\frac{2\kappa-\alpha-\gamma}{2\kappa+\alpha+\gamma}\,,\quad 
& c=\frac{\gamma-\alpha}{2\kappa+\alpha+\gamma}\,,
\\
& b=\frac{2\kappa-\delta-\beta}{2\kappa+\delta+\beta}\,, 
& d=\frac{\beta-\delta}{2\kappa+\delta+\beta}\,.
 \end{aligned}
 \right.
\end{equation}

Relations \eqref{eq:com8vertex} and \eqref{eq:bords8vertex} can be rewritten in the more usual form
\begin{equation}\begin{aligned}
 \kappa(ED-DE)=2(DH + HE), \qquad & E^2-D^2=2\kappa[H,D]=2\kappa[H,E],
\\[2ex]
 \llangle W |\big(\  \alpha E-\gamma D +2\kappa H\ \big)=0, \qquad  
 & \big(\ \delta E-\beta D -2\kappa H\ \big) | V \rrangle =0,
 \end{aligned}
\end{equation}
where $E, D$ and $H$ are the usual DEHP generators with the identification 
\be
E=G_2+G_1+G_3\mb{;} D=G_2-G_1-G_3\mb{;} 
\overline{E}=2\kappa H \mb{and} \overline{D}=-2\kappa H,
\ee
deduced from relation \eqref{eq:ide}.
In the following, we will use the natural basis  $\{G_1\,,\, G_2\,,\,G_3\}$ because it leads to simpler computations.
In particular, relations \eqref{eq:com8vertex} imply that $G_1G_3$ is central.

\subsection{Calculation of physical observables}

\paragraph{Densities.}
The average density on the site $i$ is given by
\begin{equation}
 \langle n_i \rangle = \frac{\llangle W| C^{i-1}DC^{L-i} |V\rrangle}{\llangle W| C^L |V\rrangle}
 =\frac{1}{2}\frac{\llangle W| G_2^{i-1}(-G_1+G_2-G_3)G_2^{L-i} |V\rrangle}{\llangle W| G_2^L |V\rrangle},
\end{equation}
with $C=E+D=2\,G_2$. 
We see that we  need to compute only two quantities: $\llangle W| G_2^{i-1}G_1G_2^{L-i} |V\rrangle$ and 
$\llangle W| G_2^{i-1}G_3G_2^{L-i} |V\rrangle$. For the first one, one gets
\bea \label{eq:dens1}
 \llangle W| G_2^{i-1}G_1G_2^{L-i} |V\rrangle = \phi^{i-1}  \llangle W| G_1G_2^{L-1} |V\rrangle=
\phi^{i-1} \Big( c \llangle W| G_2^{L} |V\rrangle+a  \llangle W| G_3G_2^{L-1} |V\rrangle\Big).\quad
\eea
The first equality is obtained by pushing $G_1$ to the left using the commutation relations \eqref{eq:com8vertex}. 
We then get
the second equality using the relation on the left boundary \eqref{eq:bords8vertex}.
To complete the computation, we have to deal with the last term $\llangle W| G_3G_2^{L-1} |V\rrangle$:
\begin{eqnarray} \label{eq:dens2}
 \llangle W| G_3G_2^{L-1} |V\rrangle = \phi^{L-1}  \llangle W| G_2^{L-1}G_3  |V\rrangle
 =\phi^{L-1}\Big( d  \llangle W| G_2^{L}|V\rrangle+b  \llangle W| G_2^{L-1}G_1\Big) |V\rrangle),
\end{eqnarray}
where we pushed $G_3$ to the right using \eqref{eq:com8vertex} and applied it against the right boundary using \eqref{eq:bords8vertex}.
The last term can be rewritten in the following way
\begin{equation} \label{eq:dens3}
 \llangle W| G_2^{L-1}G_1 |V\rrangle=\phi^{L-i} \llangle W| G_2^{i-1}G_1G_2^{L-i} |V\rrangle.
\end{equation}
Putting \eqref{eq:dens1}, \eqref{eq:dens2} and \eqref{eq:dens3} all together we get a closed formula for $\llangle W| G_2^{i-1}G_1G_2^{L-i} |V\rrangle$:
\begin{eqnarray}
 \llangle W| G_2^{i-1}G_1G_2^{L-i} |V\rrangle=\frac{c\phi^{i-1}+ad\phi^{L+i-2}}
 {1 -ab\phi^{2L-2}} \llangle W| G_2^{L} |V\rrangle \,.
\end{eqnarray}
In the same way we can compute $\llangle W| G_2^{i-1}G_3G_2^{L-i} |V\rrangle$:
\begin{eqnarray}
 \llangle W| G_2^{i-1}G_3G_2^{L-i} |V\rrangle= \frac{d\phi^{L-i}
 +bc\phi^{2L-i-1}}{1 -ab\phi^{2L-2}}\llangle W| G_2^{L} |V\rrangle  \,.
\end{eqnarray}
We finally get the expression of the density
\begin{eqnarray}\label{eq:density}
 \langle n_i \rangle = 
\frac{1}{2}-\frac{c\phi^{i-1}+ad\phi^{L+i-2} 
 +d\phi^{L-i}+bc\phi^{2L-i-1}}{2(1-ab\phi^{2L-2})}\:.
\end{eqnarray}
The Markov process \eqref{eq:localw8vertex} is invariant under the transformation $\kappa\rightarrow -\kappa$.
Therefore, all the physical quantities must be also invariant under this transformation. Remarking that the parameters are transformed as follows $\phi\rightarrow 1/\phi$,
$a\rightarrow 1/a$, $b\rightarrow 1/b$, $c\rightarrow -c/a$ and $d\rightarrow -d/b$, we deduce easily that density \eqref{eq:density} is indeed invariant.

\paragraph{Currents.}
The situation is here different from the previous ASEP and SSEP models, because the number of particles is not conserved in the bulk. We can define different kinds of currents: the usual current on the lattice between site $i$ and $i+1$ but also the 
evaporation current at the sites $i$ and $i+1$.
Let us first compute the usual current on the lattice from the site $i$ to $i+1$
\begin{eqnarray}
  \langle J_{i\rightarrow i+1}^{lat} \rangle &=& \frac{\llangle W| C^{i-1}\kappa^2(DE-ED)C^{L-i-1} |V\rrangle}{\llangle W| C^L |V\rrangle}
 =\frac{\kappa^2\llangle W| G_2^{i-1}(G_2G_3-G_1G_2)G_2^{L-i-1} |V\rrangle}{(\kappa+1)\llangle W| G_2^L |V\rrangle} \nonumber \\
& =& \frac{\kappa^2}{\kappa+1}\ \frac{\llangle W| G_2^{i}G_3G_2^{L-i-1} |V\rrangle}{\llangle W| G_2^L |V\rrangle}-
 \frac{\kappa^2}{\kappa+1}\ \frac{\llangle W| G_2^{i-1}G_1G_2^{L-i} |V\rrangle}{\llangle W| G_2^L |V\rrangle} \label{eq:la1}\\
& =& \frac{\kappa^2}{\kappa+1}\ \ 
 \frac{d\phi^{L-i-1}+bc\phi^{2L-i-2}-c\phi^{i-1}-ad\phi^{L+i-2}}{1-ab\phi^{2L-2}}\,.
 \label{eq:la2}
\end{eqnarray}
We can also compute the evaporation current (the evaporation is counted positively whereas the condensation is counted negatively)
\begin{eqnarray}
\langle J_{i,i+1}^{eva} \rangle &=& \frac{\llangle W| C^{i-1}(D^2-E^2)C^{L-i-1} |V\rrangle}{\llangle W| C^L |V\rrangle}
 =-\frac{\kappa}{\kappa+1}\ \frac{\llangle W| G_2^{i-1}(G_2G_3+G_1G_2)G_2^{L-i-1} |V\rrangle}{\llangle W| G_2^L |V\rrangle} \nonumber \\
& =& -\frac{\kappa}{\kappa+1}\ \frac{\llangle W| G_2^{i}G_3G_2^{L-i-1} |V\rrangle}{\llangle W| G_2^L |V\rrangle}-
 \frac{\kappa}{\kappa+1}\ \frac{\llangle W| G_2^{i-1}G_1G_2^{L-i} |V\rrangle}{\llangle W| G_2^L |V\rrangle} \label{eq:eva1} \\
& =& -\frac{\kappa}{\kappa+1}\ 
 \frac{c\phi^{i-1}+ad\phi^{L+i-2}+d\phi^{L-i-1}+bc\phi^{2L-i-2}}{1-ab\phi^{2L-2}}\,.
 \label{eq:eva2}
\end{eqnarray}
Again, it is straightforward to show that these currents are invariant under the transformation $\kappa\rightarrow -\kappa$.
We can also compute the variation of the density at the site $i$ which vanishes at the stationary state
\begin{equation}
 \frac{d \langle n_i \rangle}{dt}=0=\langle J_{i-1\rightarrow i}^{lat} \rangle-\langle J_{i\rightarrow i+1}^{lat}\rangle-\langle J_{i-1,i}^{eva} \rangle
 -\langle J_{i,i+1}^{eva} \rangle\;.
\end{equation}
This last relation can be verified algebraically with the forms \eqref{eq:la1} and \eqref{eq:eva1} or 
directly with \eqref{eq:la2} and \eqref{eq:eva2}.

\paragraph{Expressions in the large L limit.} Due to the symmetry $\phi\rightarrow 1/\phi$, we can choose $|\phi|\leq 1$. Then, from the expressions of the densities, we can deduce the following forms in the large L limit
\be\begin{array}{lll}
 \langle n_i \rangle &\displaystyle
\underset{L \to\infty}{\sim}\quad
\half\left( 1 + \frac{\alpha-\gamma}{2\kappa+\alpha+\gamma}\,\Big(\frac{\kappa-1}{\kappa+1}\Big)^\eps\right)\ ,\quad 
&i=1+\eps\,,\ \eps\geq 0
\\[2ex]
 \langle n_i \rangle &\displaystyle\underset{L \to\infty}{\sim}\quad \half\ ,\quad &i=\frac{L}2+\eps\,,
\\[2ex]
 \langle n_i \rangle &\displaystyle\underset{L \to\infty}{\sim}\quad
\half\left( 1 + \frac{\delta-\beta}{2\kappa+\delta+\beta}\,\Big(\frac{\kappa-1}{\kappa+1}\Big)^\eps\right)\ ,\quad
& i=L-\eps\,,\ \eps\geq 0
\end{array}
\ee
where $\eps$ is small. 

For $0<\phi<1$ (i.e. $1<\kappa$), it shows that depending on the sign of the difference $\alpha-\gamma$ (resp. $\delta-\beta$), the density closed to the left (resp. right) boundary 
is larger or smaller than the average density in the bulk. 

\begin{figure}[htbp]
\begin{center}
\includegraphics[width=160mm,height=140mm]{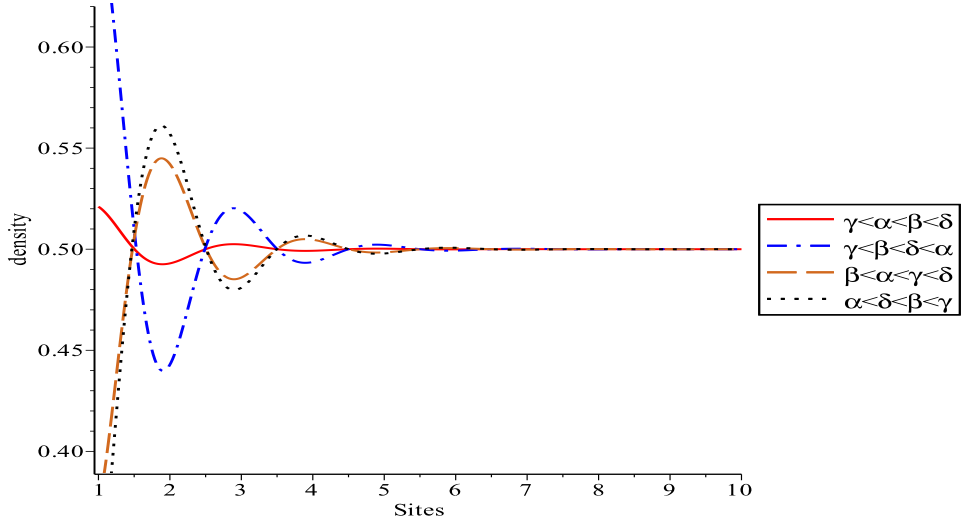}
\end{center}
\vspace{-60mm}
\caption{Density profiles close to the left boundary for various values of the boundary parameters\label{fig:densites}}
\end{figure}

For $-1<\phi<0$ (i.e. $0<\kappa<1$) and close to the boundaries, the density shows Friedel oscillations  (see figure \ref{fig:densites}).

In the same way, we get for the lattice current
\be\begin{array}{lll}
\langle J_{i\rightarrow i+1}^{lat} \rangle &\displaystyle
\underset{L \to\infty}{\sim}\quad
\frac{\kappa^2}{1+\kappa}\ \frac{\alpha-\gamma}{2\kappa+\alpha+\gamma}\,\Big(\frac{\kappa-1}{\kappa+1}\Big)^\eps
\ ,\quad 
&i=1+\eps\,,\ \eps>0
\\[2ex]
\langle J_{i\rightarrow i+1}^{lat} \rangle &\displaystyle
\underset{L \to\infty}{\sim}\quad 0\ ,\quad &i=\frac{L}2+\eps\,,
\\[2ex]
\langle J_{i\rightarrow i+1}^{lat} \rangle &\displaystyle
\underset{L \to\infty}{\sim}\quad
-\frac{\kappa^2}{1+\kappa}\ \frac{\delta-\beta}{2\kappa+\delta+\beta}\,\Big(\frac{\kappa-1}{\kappa+1}\Big)^{\eps-1}\ ,\quad
& i=L-\eps\,,\ \eps>0
\end{array}
\ee
and for the evaporation current
\be\begin{array}{lll}
\langle J_{i,i+1}^{eva} \rangle &\displaystyle
\underset{L \to\infty}{\sim}\quad
\frac{\kappa}{1+\kappa}\ \frac{\alpha-\gamma}{2\kappa+\alpha+\gamma}\,\Big(\frac{\kappa-1}{\kappa+1}\Big)^\eps
\ ,\quad 
&i=1+\eps\,,\ \eps>0
\\[2ex]
\langle J_{i,i+1}^{eva} \rangle &\displaystyle
\underset{L \to\infty}{\sim}\quad 0\ ,\quad &i=\frac{L}2+\eps\,,
\\[2ex]
\langle J_{i,i+1}^{eva} \rangle&\displaystyle
\underset{L \to\infty}{\sim}\quad
\frac{\kappa}{1+\kappa}\ \frac{\delta-\beta}{2\kappa+\delta+\beta}\,\Big(\frac{\kappa-1}{\kappa+1}\Big)^{\eps-1}\ ,\quad
& i=L-\eps\,,\ \eps>0\,.
\end{array}
\ee
As for densities, when $0<\phi<1$, the sign of the current close to the left (resp. right)
boundary depends on the sign of the difference $\alpha-\gamma$ (resp. $\delta-\beta$), while for $\phi<0$ it oscillates.

\subsection{Representation of the algebra and the boundary vectors}

In this section we give an explicit representation of the generators $G_1$, $G_2$, $G_3$ and of the boundary vectors $|V \rrangle$ and $\llangle W|$.
It proves the existence of these objects and the fact that the normalization factor $\llangle W|G_2^L |V\rrangle$ appearing in the previous
algebraic computations is not equal to zero.

Let us set
\begin{equation}
 G_1=g_1 \otimes 1, \quad G_2=g_2 \otimes g_2, \quad G_3=1 \otimes g_3,
\end{equation}
where
\begin{equation}
 g_1=\sum_{n=0}^{+\infty}|n+1\rrangle \llangle n|, \quad g_2=\sum_{n=0}^{+\infty}\phi^n|n\rrangle \llangle n|, \quad g_3=\sum_{n=1}^{+\infty}|n-1\rrangle \llangle n|.
\end{equation}
We have used  $\{|n\rrangle\ |\ n\geq 0\}$ as an infinite basis of the additional space. 
One can check that these infinite matrices verify the commutation relations \eqref{eq:com8vertex}.

Let us write $|V\rrangle$ in the following way
\begin{equation}
 |V\rrangle =\sum_{n,m=0}^{+\infty}v_{n,m}\,|n\rrangle \otimes |m\rrangle.
\end{equation}
The equation on the right boundary \eqref{eq:bords8vertex} is then equivalent to the equations
\begin{equation} 
 \left\{ \begin{aligned}
          & -bv_{n-1,m}-d\phi^n\phi^m v_{n,m}+v_{n,m+1}=0, \quad n\geq 1, \ m\geq 0, \\
          & -d\phi^m v_{0,m}+v_{0,m+1}=0, \quad m\geq 0.
         \end{aligned}
 \right.
\end{equation}
It is easy to check through a direct computation that
\begin{equation}
v_{n,m}= d^{m-n}b^n \frac{\phi^{\frac{(m-n)(m-n-1)}{2}}}{(1-\phi^2)\dots (1-\phi^{2n})}
\end{equation}
is a solution to these equations.

In the same way for the other boundary, we define
\begin{equation}
 \llangle W|=\sum_{n,m=0}^{+\infty} w_{n,m}\, \llangle n| \otimes \llangle m|,
\end{equation}
and the equation \eqref{eq:bords8vertex} on the left boundary is equivalent to the equations
\begin{equation} 
 \left\{ \begin{aligned}
          & -aw_{n,m-1}-c\phi^n\phi^m w_{n,m}+w_{n+1,m}=0, \quad m\geq 1, \ n\geq 0, \\
          & -c\phi^n w_{n,0}+w_{n+1,0}=0, \quad n\geq 0.
         \end{aligned}
 \right.
\end{equation}
One can check without any difficulty that
\begin{equation}
w_{n,m}= c^{n-m}a^m \frac{\phi^{\frac{(n-m)(n-m-1)}{2}}}{(1-\phi^2)\dots (1-\phi^{2m})}
\end{equation}
is a solution to these equations.

One can compute the normalization factor
\begin{equation}
 \llangle W| G_2^L |V\rrangle = \sum_{n,m=0}^{+\infty} \left(\frac{cb\phi^L}{d}\right)^n \left(\frac{da\phi^L}{c}\right)^m
 \frac{\phi^{(m-n)^2}}{(1-\phi^2)\dots (1-\phi^{2n})\times (1-\phi^2)\dots (1-\phi^{2m})}.
\end{equation}
Using the Stolz-Cesaro criteria \cite{stolz}, one can show that the series is convergent when 
\be\label{convergence}
|\phi|<1\,, \quad
\left|\frac{b\,c\,\phi^L}{(1-\phi^2)d}\right|<1 \mb{and} \left|\frac{a\,d\,\phi^L}{(1-\phi^2)c}\right|<1.
\ee
Remark that to thanks to the symmetry $\kappa\rightarrow -\kappa$, we can choose $|\phi|<1$. Then, when $L$ is large enough,
the conditions \eqref{convergence} are always fulfilled (when $c,d\neq0$). 
 
\section{Conclusion}
In this paper we analyzed the matrix ansatz in the integrable systems framework. We showed that the quadratic algebra arising with the matrix ansatz can be linked in some cases to quantum groups. This connection is made in two steps. First, when the R-matrix of the model satisfies the Markovian property, it is possible to express linearly generators of the Zamolodchikov algebra in terms of generators of the quantum group.
Second the ZF relation gives back, through a derivation, the bulk relation of the matrix ansatz.

We gave the explicit expression, in the out-of-equilibrium statistical physics language, of 
the key objects (R and K matrices) needed for the integrable techniques. As a byproduct, 
we proved the integrability of the open TASEP model. We constructed the transfer matrix and 
proved, for the SSEP and ASEP models, that the vector given by the matrix ansatz is an eigenvector 
of the transfer matrix. We computed the associated eigenvalue and connected it to the 
algebraic Bethe ansatz framework. It would be interesting to push further this investigation and understand 
if this eigenvector could be used as a reference state (i.e. a highest/lowest weight for the monodromy matrix) 
for the algebraic Bethe ansatz. Another fruitful way to tackle this problem would be to generalize the coordinate matrix ansatz developed in \cite{CRS1}
allowing one to get other eigenvalues and eigenvectors.

In order to show the efficiency of the method, we constructed an algebra that allows us to write in a matrix product form the stationary state of a 
reaction-diffusion model. We computed physical observables using the algebraic relations in the bulk and on the boundaries.
We believe that this approach is quite general: once the R-matrix of an integrable system is known, it is possible to compute the K matrices 
by solving the reflection equation. Then the ZF relation gives the exchange relations of the generators in the bulk and the GZ relation 
gives the action of the generators on the boundaries. This method can be for instance used to construct the quadratic algebra which 
allows us to compute the weight of the stationary state of a 2-species TASEP model with open boundaries. This algebra contains more 
than three generators and gives non scalars overlined operators in the matrix ansatz. We will come back on this case in a forthcoming publication.

The choice of an evaluation map for a given problem is also an open question. Indeed, the level of truncation of the series $A(x)$ was 
found case by case and through tries and errors for the models we presented. A better understanding of 
 the optimal number of generators to describe a stationary state in a given model is a crucial point in the method.

Another interesting development would be to prove the existence of non zero boundary vectors $\llangle W|$ and  $|V\rrangle$ satisfying 
the GZ relations without constructing case by case explicit representations. We believe that the ZF and  GZ relations should offer 
efficient ways to compute physical observables. 
Finally the matrix ansatz only works for a Markovian Hamiltonian, it could be interesting to generalize it to a current counting 
deformation of the Hamiltonian (this deformation is still integrable in the SSEP, ASEP and TASEP models). Work has already be done 
in this direction \cite{GLMV,laza,LP}.

\paragraph{Acknowledgement:} M. Vanicat thanks the Laboratoire Charles Coulomb for hospitality and financial support during his stay.
We thank L. Cantini, J. de Gier, K. Mallick and V. Rittenberg for their interests and their suggestions improving the version 2 of this paper.

\end{document}